\documentclass[12pt]{article}
\usepackage[margin=1in]{geometry}
\usepackage{fullpage}
\usepackage{amsmath}
\usepackage{amsfonts}
\usepackage{amssymb}
\usepackage{amsthm}
\usepackage{graphicx}

\usepackage{rotate,rotating}
\usepackage{setspace}

\usepackage{geometry}
\usepackage{enumerate}
\usepackage{epstopdf}
\usepackage{enumitem}
\usepackage{pdfpages,xcolor}
\usepackage{float}
\usepackage{subcaption}
\definecolor{navyblue}{rgb}{0,0,0.8} 
\usepackage[colorlinks, linkcolor=navyblue, citecolor=navyblue, urlcolor=navyblue, pdfborder={0 0 1}, citebordercolor={1 1 1}]{hyperref}
\usepackage{authblk}

\usepackage{booktabs}
\usepackage{array}
\usepackage{ulem}
\usepackage{tikz}
\usepackage{natbib}
\usepackage{appendix}

\usepackage{threeparttable} 

\usepackage[utf8]{inputenc} 
\usepackage[T1]{fontenc}    
\usepackage{hyperref}       
\usepackage{url}            
\usepackage{nicefrac}       
\usepackage{microtype}      

\usetikzlibrary{shapes.geometric}

\usepackage{array} 
\usepackage{tabularx}

\usepackage{cleveref}

\usepackage{fancyhdr}  

\title{\textbf{Forecasting Intraday Volume in Equity Markets 
\\with Machine Learning}
}

\author[1,2,3]{Mihai Cucuringu}
\author[2]{Kang Li}
\author[4]{Chao Zhang}

\affil[1]{\small Department of Mathematics, University of California Los Angeles, US}
\affil[2]{\small Department of Statistics, University of Oxford, Oxford, UK}
\affil[3]{\small Oxford-Man Institute of Quantitative Finance, University of Oxford, Oxford, UK}
\affil[4]{\small FinTech Thrust, HKUST(GZ), Guangzhou, China}

\date{First Version: April 2025}

\newboolean{showComments}
\setboolean{showComments}{false}

\newboolean{showColors}
\setboolean{showColors}{false}

\definecolor{colour1}{RGB}{166,206,227}  
\definecolor{colour2}{RGB}{31,120,180} 
\definecolor{colour3}{RGB}{178,55,250} 
\definecolor{colour4}{RGB}{51,160,44} 
\definecolor{colour5}{RGB}{158,155,50} 

\ifthenelse{\boolean{showComments}}{
    \newcounter{noteMCctr} \setcounter{noteMCctr}{1}
    \newcommand{\MC}[1]{\textbf{\textcolor{colour3}{{{Mihai \#\arabic{noteMCctr}:}}#1}} \addtocounter{noteMCctr}{1}}

    \newcounter{noteXXctr} \setcounter{noteXXctr}{1}
    \newcommand{\KL}[1]{\textbf{\textcolor{red}{{{KL: \#\arabic{noteXXctr}: }}#1}} \addtocounter{noteXXctr}{1}}

    \newcounter{noteZZctr} \setcounter{noteZZctr}{1}
    \newcommand{\CZ}[1]{\textbf{\textcolor{blue}{{{CZ: \#\arabic{noteZZctr}: }}#1}} \addtocounter{noteZZctr}{1}}
}{
    \newcommand{\CZ}[1]{{}}
    \newcommand{\KL}[1]{{}}
    \newcommand{\MC}[1]{{}}
}

\ifthenelse{\boolean{showColors}}{
    \newcommand{\mc}[1]{\textcolor{colour3}{#1}}
    
    \newcommand{\kl}[1]{\textcolor{red}{#1}}
 }{
    \newcommand{\mc}[1]{\textcolor{black}{#1}}
    
    \newcommand{\kl}[1]{\textcolor{black}{#1}}
 }

\begin{document}

\maketitle

\def\thefootnote{\arabic{footnote}}

\begin{abstract}

This study focuses on forecasting intraday trading volumes, a crucial component for portfolio implementation, especially in high-frequency (HF) trading environments. 
Given the current scarcity of flexible methods in this area, we employ a suite of machine learning (ML) models enriched with numerous HF predictors to enhance the predictability of intraday trading volumes. 
Our findings reveal that intraday stock trading volume is highly predictable, especially with ML and considering commonality.
Additionally, we assess the economic benefits of accurate volume forecasting through Volume Weighted Average Price (VWAP) strategies. 
The results demonstrate that precise intraday forecasting offers substantial advantages, providing valuable insights for traders to optimize their strategies.

\end{abstract}

{
}

\newpage

\section{Introduction} \label{sec:intro}

Extensive research on portfolio construction has been devoted to return prediction and, albeit to a lesser extent, covariance prediction. While the implications of mean return and covariance forecasts are thoroughly examined in the literature, how to effectively implement \mc{and deploy} these portfolios remains underexplored, posing significant challenges in practice.\footnote{\mc{A strand of literature has focused on trading cost analysis}, including  \cite{kyle_continuous_1985, garleanu_dynamic_2016, frazzini_trading_2018}.} 

In order to address this gap, understanding the dynamics of trading volumes becomes crucial, as they directly impact the portfolio execution. Generally speaking, trading costs increase as the partition rate -- defined as the ratio of trade size relative to market trading volume -- rises, and decrease as it falls \citep{kyle_continuous_1985}. In other words, higher market trading volumes allow for more aggressive trading, concealing trade intentions, whereas lower volumes increase costs and risk revealing strategic trades, as shown in \cite{goyenko2024trading}.

Furthermore, accurate intraday volume forecasts enhance optimal order scheduling, namely, how to split a large parent order into smaller child orders to secure better price execution and minimize costs \citep{bankOptimalOrderScheduling2013}. On the other hand, such forecasts improve the replication of the VWAP strategy \citep{dynamicVolume}, a common benchmark in practice to evaluate the performance of trades in terms of their ability to execute orders.\footnote{VWAP is defined at the end of the trading day as an average of intra-daily transaction prices weighted by the corresponding share of traded volume relative to the total daily volume.}

Despite its importance, forecasting intraday trading volumes has received relatively little attention in the literature. One notable study is the Component Multiplicative Error Model (CMEM) by \cite{Brownlees_CMEM}, which effectively captures intraday trading volume dynamics. CMEM decomposes these dynamics into three components: (i) a daily component capturing day-to-day volume trends, (ii) an intraday periodic component aiming to reproduce the time-of-day pattern, and (iii) an intraday non-periodic component. 

However, CMEM has several limitations that our research aims to overcome. As \cite{kf_cmem2016} noted, the model's estimation accuracy may be influenced by initial parameter settings and the positivity conditions on each component. Furthermore, while CMEM excels in univariate time series modeling for intraday trading volumes, it lacks the capability to incorporate cross-sectional information, a critical dimension as studies such as \cite{chordia2000commonality, brockman2002commonality} demonstrated commonality in liquidity across markets and industries. Lastly, the advent of high-frequency trading data calls for a more flexible model capable of leveraging richer information and accounting for potential nonlinear relationships.

In recent years, machine learning (ML) models, including deep neural networks (NN), have achieved substantial advancements in fields such as computer vision and natural language processing, and have shown great potential in finance, particularly in tasks of modeling returns and volatilities.  For example, \cite{gu2020empirical, Kolm2021DeepOF} demonstrated the efficacy of ML approaches in forecasting returns, due to their ability to handle a large set of variables and their nonlinear relations. \cite{zhang2024volatility} employed a suite of ML models to forecast intraday realized volatility and found that NNs yield superior out-of-sample forecasts over a strong set of traditional baselines. However, the application of ML in forecasting intraday trading volumes remains less explored, as noted by \cite{goyenko2024trading}.

In the present study, we investigate a variety of ML models for forecasting multi-asset intraday trading volumes by leveraging high-frequency data from the U.S. equity market. We first enhance the basic CMEM model with a set of {auxiliary predictors}, such as the number of trades, derived from high-frequency limit order book data from LOBSTER \citep{huang_lobster_2011}. Our findings indicate that NNs equipped with an extended set of predictors significantly outperform benchmark models in out-of-sample tests.

To leverage the commonality in intraday trading volumes, we propose a strategy that first clusters assets and then trains a specialized model or ``expert'' for each cluster. This strategy spans from training individual models for each asset to developing a universal model for all considered assets. The results reveal that taking advantage of this commonality leads to improved prediction performance. Moreover, nonlinear models, including ensemble trees and NNs provide better predictability over simple linear models, when controlling for the same set of variables and information content.

Finally, we assess the economic benefits of the enhanced intraday volume forecasts through the application of a VWAP strategy. Our evaluation first focuses on the VWAP replication error, where lower values indicate a closer match to the VWAP and imply minimal market impact. Additionally, we employ a matching engine, as detailed by \cite{frey_jax-lob_2023}, to examine the fill ratio of passive orders.\footnote{It is important to note that this matching engine uses historical order replay, providing a more realistic assessment than hypothetical simulations.}

The remainder of this paper is organized as follows. We begin with \Cref{sec:literature} by reviewing related literature. 
\Cref{sec:data_featrues} describes our dataset, defines the variables of interest, and exhibits intraday volume patterns. 
In \Cref{sec:Methodology}, we introduce the benchmark CMEM, three training schemes, and various machine learning models for predicting future intraday volumes. 
\Cref{sec:methods} provides the forecasting results and offers a discussion of their implications.
\Cref{sec:application} explores the application of these forecasts in VWAP strategies, and \Cref{sec:conclusion} concludes the paper with a summary and an outlook on future work.

\section{Literature Review} \label{sec:literature}

Our study is built on several research streams, such as the cross-sectional commonality, the decomposition of traded volume, the application of ML models, and \mc{and employing an expanded set of} predictors. 
To begin with, the cross-sectional commonality is explored in the existing literature on the topic of forecasting intraday volume. In the Additive Error Model proposed by \cite{dynamicVolume}, the \textbf{cross-sectional} data, representing information about all stocks at a specific point in time, and the \textbf{time series} data, capturing the volume changes for each stock throughout the trading day, are used simultaneously to estimate the model parameters. In this research, a matrix with columns \mc{corresponding to different stocks and rows to different bins of volume data is used as input to a method that} predicts the volume of all stocks. 
This is different from our proposed methodology because \cite{dynamicVolume} only utilizes volume data, while we consider \mc{a larger number of} predictors and discover the commonality 
The second stream is about the decomposition of intraday volumes to better understand the endogenous properties and seasonality of volume from a time series perspective. 
A noteworthy contribution to this stream is the model proposed by \cite{dynamicVolume}, which employed an additive decomposition to articulate the intraday U-shape while adeptly integrating both periodic patterns and autocorrelations of intraday volumes. Similarly, the Component Multiplicative Error Model (CMEM), as introduced by \cite{Brownlees_CMEM}, was built upon the Multiplicative Error Model.\footnote{It was originally put forward by \cite{engle2002} and further elaborated by \cite{engle2006}.} The intraday volume was decomposed into three components, namely, a daily average component, an intraday periodic component, and an intraday dynamic component. This approach outperformed the Rolling Mean\footnote{The Rolling Mean approach refers to taking the rolling mean of a specific bin from the past several days as the forecasted value of the intraday volume for that same bin on the current day.} approach on ETFs. A comparative analysis of the aforementioned multiplicative model from \cite{Brownlees_CMEM} and the additive model from \cite{dynamicVolume} was undertaken in \cite{two_early_models}. In the comparison, the additive model was found to \mc{perform better} as it incorporates not only time series data but also cross-sectional information. 

Based on the decomposition of the CMEM model, \cite{kf_cmem2016} transformed the multiplicative framework into an additive one through logarithmic conversion, given the right skewness \footnote{Right skewness indicates that the distribution has a long tail on the right side, with a few observations being much larger than the majority. While the logarithmic transformation compresses the range of values, reducing the impact of extreme outliers and making the distribution more symmetrical.} in the intraday volume distribution observed by \cite{volume_skewness}. This approach further accommodated the enhancement of robustness by applying LASSO regularization to effectively manage anomalies. The model was then estimated using the Kalman filter from a state-space model perspective. 

Aside from the multiplicative model, several other research studies have also explored the decomposition of intraday volume as a univariate time series, applying time series analysis techniques to break down its components. \cite{volumePercentage} applied an ARMA (Auto-Regressive Moving Average) model to predict the components. Moreover, the work of \cite{wavelet2010} employed wavelet decomposition to dissect historical intraday volume data. \cite{arneric_multiple_2021} employed multiple STL decompositions to discover the seasonal and trend decomposition of intraday volume.

Furthermore, machine learning models have demonstrated great potential in financial applications, and this line of work has also proven to be effective in the context of intraday volume forecasting. 
\cite{svm2015} proposed that the intraday volume was decomposed into systematic patterns and residual components, demonstrating that the former can be projected using the moving average of preceding days, whereas the residuals were more effectively predicted by \kl{the Support Vector Machine (SVM).} 
Other ML predictive techniques, such as tree-based models and temporal mixture ensemble models, have found applications in forecasting the intraday trade volume of cryptocurrencies, \kl{as discussed by \cite{antulov-fantulin_temporal_2021}.}

Nonetheless, little attention has been paid to expanding the set of predictors in order to enhance intraday volume predictions.  \cite{Kepler2021} incorporated additional predictors, such as price volatility, the logarithm of trade size, and the log of the daily trade count, to predict intraday volume. Additionally, \cite{peikingU2022} delved into higher-frequency periodicities by examining intervals ranging from 1 to 5 minutes, thereby extending beyond the characteristic U-shaped intraday volume pattern.

Due to the application of various prediction methods, there has been a significant improvement in the accuracy of intraday volume predictions. This has led to promising applications, with one of the most common being VWAP strategies. Widespread in industry, VWAP strategies aim to execute trades at or near the Volume Weighted Average Price over a specified time horizon and constitute a considerable segment of total institutional trading. As discussed by \cite{VWAPStrategiesPortfolio} and \cite{kissellPracticalFrameworkEstimating2004}, these strategies rely on forecasts of intraday trading volumes, and their effectiveness is closely tied to the accuracy of these volume predictions. For instance, \cite{volumeDistributions} and \cite{kf_cmem2016} have observed a reduction in VWAP tracking error\footnote{The VWAP tracking error is a measure of the effectiveness of VWAP strategies. This will be discussed in detail in \Cref{sec:application}.} with more accurate volume prediction models.

\section{Data and Variables} \label{sec:data_featrues}

\subsection{Dataset Description}

In our analysis, we employ constituents of the S\&P 500 index from July 2017 to December 2017. The data calculated from the LOBSTER dataset proposed by \cite{huang_lobster_2011}, which contains the best 10 price levels. We restrict our analysis to a universe of 469 stocks, including only those that traded continuously from the start to the end of our sample period. To align with the frequency studied in the CMEM model (see \cite{Brownlees_CMEM}), we divide each day into 15-minute intervals, to which we refer as bins. We calculate the intraday volume and predictors based on the data within each bin. The volume used in our analysis is the traded volume, expressed in the number of shares\footnote{To represent traded volume, we use the sum of the \textbf{Number of shares in buy orders} and the \textbf{Number of shares in sell orders}.}. 

\subsection{Intraday Trading Pattern}

\begin{figure}[htbp]
\centering
\includegraphics[width=0.75\linewidth]{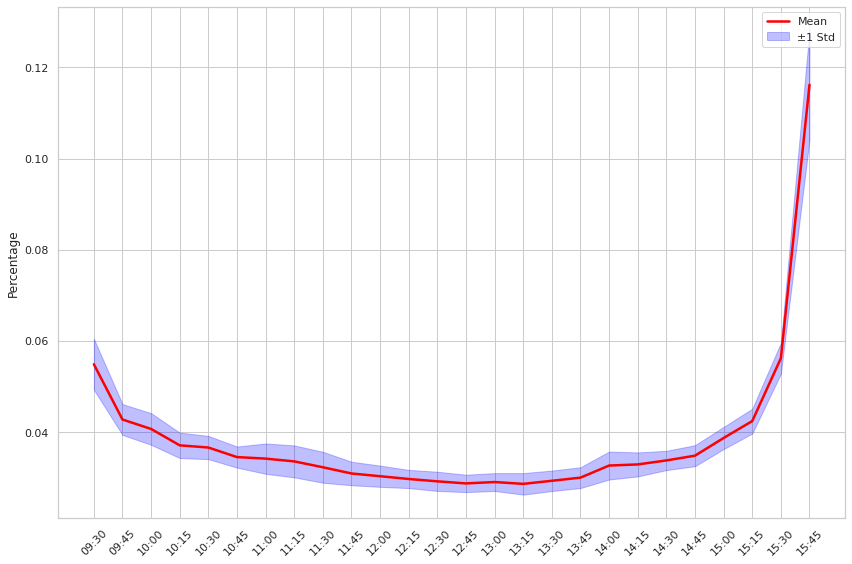}
\caption{\textbf{Intraday Trading Volume.} The plot shows the fluctuations in trading volume over a trading day period. The solid line denotes the mean, and the shaded area denotes one standard deviation. The value has been normalized by dividing it by the mean of one day's trading volume.}   
\label{fig:imageUshape}
\end{figure}

Previous works from the extant literature have documented the U-shape or J-shape pattern in the intraday trading volume. For example, \cite{volumePatternTiming} identified an intraday volume pattern described as reverse L-shaped, also commonly referred to as the J-shaped pattern. This pattern is characterized by a high trading volume at the market open and close, with a dip in the middle of the trading day. Many factors may contribute to this pattern. \cite{volumeOF} observed that intraday trading volume, along with orders from both informed and uninformed traders, followed a J-shaped pattern. Furthermore, both informational and liquidity-driven trading activities contribute to explaining the intraday dynamics of trading volume. \cite{min2018crosssectional} linked the pattern in trading volume to the execution of mutual fund inflows and outflows, which were benchmarked to the day's closing price. In addition, \cite{volumePatternETF} suggested that the observed trading volume pattern could be attributed to information-based trading, resulting in high volumes at the market open, and portfolio rebalancing activities, leading to increased volumes towards the market close.

In our research, we also observe that the intraday volume during the day has a  J-shape or U-shape (see \Cref{fig:imageUshape}).  The last bin, due to including the closing auction volume, exhibits a high intraday volume.\footnote{Here are several explanations for the high volume in the final bin. In addition to the typical orders receive throughout the day, two new orders are received during the final bin, which have contributed to the observed increase in intraday volume. These two orders have been mentioned in \cite{nasdaq2020openclose}, including \textit{Limit-on-Close (LOC) orders} received after 3:55 p.m. and the acceptance of \textit{On-Close Orders}. Similarly, as stated in \cite{nyse2023holidays}, the period from 3:50 p.m. to 4:00 p.m. is designated as the closing imbalance period, which can also impact the intraday volume in the final bin.} 

A structured model is used in our research to characterize this pattern. In addition to the intraday volume pattern, the autocorrelation of  the intraday volume is also analyzed in the \Cref{sec:autocorrelation}

\subsection{Feature Description and Feature Engineering} \label{sec:fetures}

\kl{Building upon our observations on the intraday trading volume patterns, we develop a feature engineering framework to effectively capture both the temporal dynamics and multiple dimensions of market activity for volume prediction. We categorize our features into \textit{basic predictors} derived directly from market data,  and \textit{compound predictors} that incorporate temporal relationships. This structured methodology enables our models to leverage both contemporaneous market conditions and historical trading patterns, which is essential for accurate intraday volume forecasting. The mathematical relationship between our basic and compound predictors is formalized as}

\begin{equation}
\begin{aligned}
\text{ Compound Predictor }  = \text{Operation } \otimes \text{ Basic Predictor }
\end{aligned}
\label{eq:compound_feature}
\end{equation}

\Cref{tab:basic_feature} lists the basic predictors for our forecasting task. It is worth noting that the indicator \texttt{timeHMs} is converted into a categorical variable, which is effective in predicting the intraday volume due to the U-shape pattern of the volume throughout the day. Given that market activity is typically more intense during the opening and closing periods, with potentially different market dynamics compared to other trading hours, we define the open interval 9:30-10:00, mid-day interval 10:00-14:30, and close interval 14:30-15:00. We then incorporate the intraday interval as an additional predictor alongside \texttt{timeHMs}, providing a less granular temporal feature that enhances our predictive capability. Furthermore, we consider volume from three different perspectives: trading notional amount, number of shares traded, and number of trades. Incorporating these different dimensions as features into the model improves the prediction accuracy of the intraday volume, as illustrated by the feature importance ranking analysis in Table \ref{tab:feature_importance}.

\kl{Due to the autocorrelation exhibited by trading volume and its predictors in time series, we need to consider both cross-sectional and temporal data dimensions. Our initial predictors were primarily derived from cross-sectional data. However, to capture temporal information, we integrate these basic predictors across the time series, thereby synthesizing new compound predictors for our models, as demonstrated in \eqref{eq}. For instance, the trading volume during the opening interval on day $t+1$ correlates with the total volume on day $t$, a temporal dependency that requires time series features to capture effectively. Specifically, when applying the operation \texttt{past\_2} to the basic predictor \texttt{nrTrades}, we generate the compound predictor \texttt{nrTrades\_2}, representing the sum of trades over the previous two intervals (i.e., the preceding 30 minutes). To systematically develop these compound predictors, we employ a set of predefined operations. \Cref{tab:compound_features} provides a comprehensive list of all the operations applied to the basic predictors. Notably, operations \texttt{past\_2} and \texttt{past\_8} are chosen according to the ACF analysis, which revealed stronger autocorrelations at these specific lag periods.}

\begin{table}[H]
\caption{\textbf{The basic predictors for forecasting} These predictors are statistical characteristics computed from LOBSTER data within 15-minute basic bins. }
\centering
\begin{tabular}{ll}
\toprule
\textbf{Basic Predictors} & \textbf{Description} \\
\hline
\texttt{timeHMs} & Start time (hours, minutes) \\
\texttt{intrIn} & Intraday interval (open, mid-day, close) \\
\texttt{volBuyNotional} &  Buy notional value in USD \\
\texttt{volSellNotional} &  Sell notional value in USD \\
\texttt{volBuyNrTrades\_lit} & Number of trades for buy visible(lit) orders   \\
\texttt{volSellNrTrades\_lit} & Number of trades for sell visible(lit) orders \\
\texttt{volBuyQty} & Number of shares in buy orders \\
\texttt{volSellQty} & Number of shares in sell orders \\
\texttt{nrTrades} & Number of trades executed \\
\texttt{ntr} & Number of trades, which comes from lit orders \\
\bottomrule
\end{tabular}
\label{tab:basic_feature}
\end{table}

\begin{table}[H]
\caption{\textbf{Ways of synthesizing compound predictors.} We synthesize compound predictors by applying these operations to basic predictors.}
\centering
\begin{threeparttable}
\begin{tabular}{ll}
\toprule
\textbf{Operations} & \textbf{Description} \\
\midrule
\texttt{daily} & Sum the data across all bins within a day \\
\midrule
\texttt{intraday}  & Sum within each group: opening, mid-day, and closing \text{intervals}. \\
\midrule
\texttt{\text{past\_2}} & Sum up statistics from the previous two bins, \\
&covering data from the last 30 mins. \\
\midrule
\texttt{\text{past\_8}} &  Sum up statistics from the previous eight bins, \\
&covering data from the last 2 hours. \\
\bottomrule
\end{tabular}

\end{threeparttable}
\label{tab:compound_features}
\end{table}

\section{Methodology} \label{sec:Methodology}
\subsection{The Benchmark: CMEM}

The CMEM is the benchmark in the extant literature for the task of forecasting intraday trading volume. \kl{This model, introduced by \cite{Brownlees_CMEM}, builds upon the Multiplicative Error Model framework originally developed by \cite{engle2002} and further elaborated by \cite{engle2006}.} It is designed to capture the U-Shape pattern in intraday trading volumes with a multiplicative error structure, where $\epsilon_{tj}$ is an iid error term

\begin{equation}
    x_{tj} = \eta_t \cdot s_j \cdot \mu_{tj} \cdot \epsilon_{tj}.
\label{eq:cmem}
\end{equation}

This model decomposes the volume into three components:
\begin{itemize}
\item \textbf{A daily component} $\eta_t$ to model the overall level or mean of volumes at day $t$. This follows an autoregressive structure to capture day-to-day persistence.

\item \textbf{An intraday periodic component} $s_j$ to represent periodic time-of-day seasonalities, also known as the intraday U-pattern.

\item \textbf{An intraday non-periodic component} $\mu_{tj}$ to capture intraday autocorrelations beyond periodicity.

\end{itemize}
By employing the Generalized Method of Moments, these parameters can be estimated jointly. \kl{The method proceeds by first decomposing the volume data into daily, periodic, and non-periodic components, then constructing moment conditions based on the standardized residuals between actual and predicted volumes, and then extracting efficient variable instruments from these conditions using the model's gradient structure. Finally, one then solves the resulting system of equations through numerical methods that iteratively update parameter estimates until the weighted sum of these moment conditions converges to zero, thereby capturing the dynamics across all three components of the model.}

\subsection{Training Scheme}
As pointed out by  \cite{min2018crosssectional},  volume forecasts are correlated across stocks. Inspired by \cite{dynamicVolume}, where the cross-sectional relationship was used for prediction, we follow a similar procedure to utilize the commonality in the intraday volume. Specifically, the following three training schemes, which are proposed by \cite{emmenual},  are explored in our model

\begin{itemize}
\item \textbf{The Single Asset Model (SAM)} delineates the approach whereby each trading volume forecast is formulated utilizing the individual predictors of a stock.

\item \textbf{The Clustered Asset Model (CAM)} describes a methodology in which the predictors from all members of a cluster, or ‘neighbors’, are integrated into the models. This method enhances the models by incorporating shared characteristics through the aggregation of neighbors' predictors. For a detailed description of the implementation of CAM in our research, please refer to \Cref{sec:cam_detials}.

\item \textbf{The Universal Asset Model (UAM)} trains a single model on the pooled data of all stocks in
our dataset. As emphasized by \cite{sirignanoUniversalFeaturesPrice2018}, a model trained on pooled data provided better performance in predicting the direction of price movements compared to models trained on the time series of individual stocks. Under the task of forecasting intraday volume, we achieve improved model performance by identifying and leveraging the commonalities across stocks.
\end{itemize}

\subsubsection{Clustered Asset Model} \label{sec:cam_detials}
\begin{figure}[htbp]
    \centering
    \includegraphics[width=0.65\linewidth]{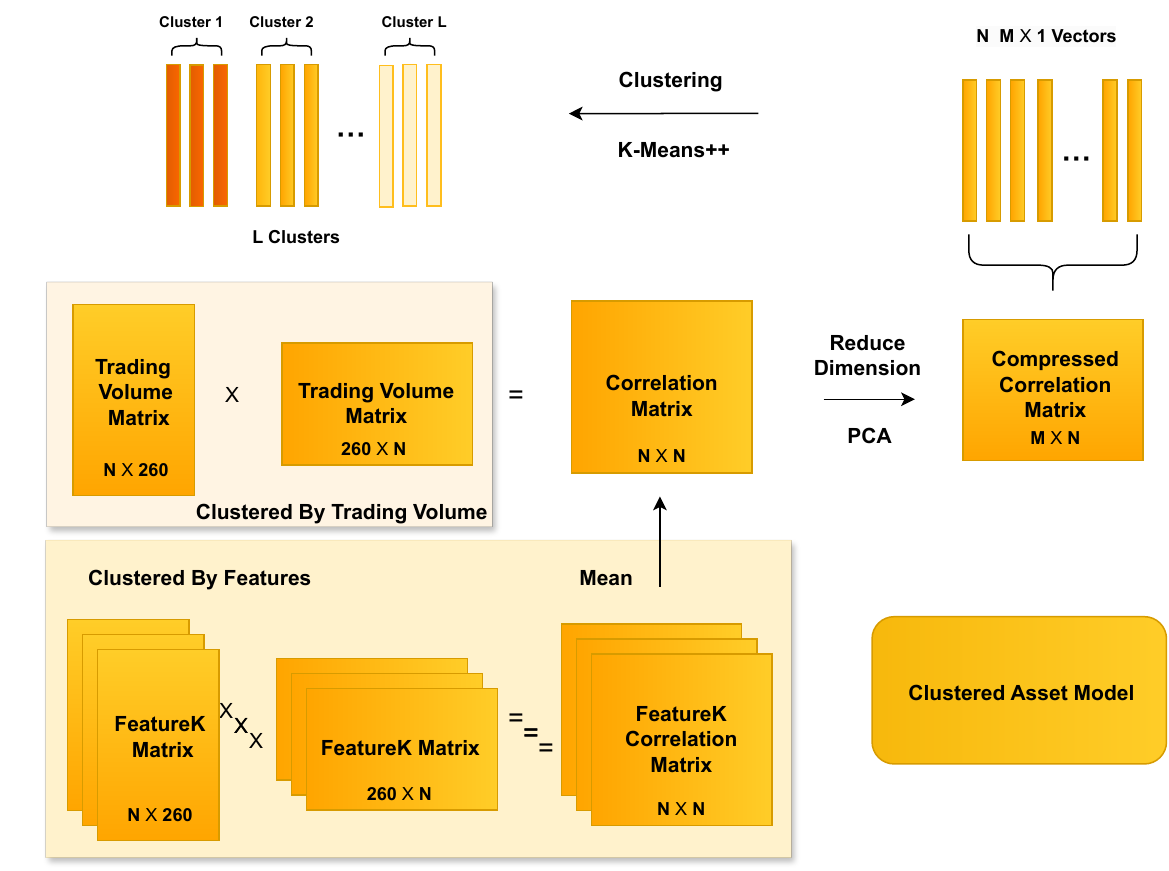}
    \caption{\textbf{Clustered Asset Model Diagram}. Two inputs are used to obtain correlations: trading volume and features. After generating the correlation matrix by clustering, PCA is employed to compress the information from the matrix. Subsequently, the K-Means++ algorithm is applied for clustering, dividing stocks into multiple groups.}
    \label{fig:diagram_clustered}
\end{figure}

We follow two clustering approaches in our research: \textbf{data clustered by trading volume} and \textbf{data clustering by features}\footnote{Related experiments on \textit{data clustering by features} have been conducted, and the results indicate that the experimental results of these two approaches are similar. For a more detailed discussion, please refer to the \Cref{sec:comparison_CAM}. }, which corresponds to performing the clustering based on the output and input for the forecasting task, respectively. For the \textbf{data clustered by trading volume}, firstly, we compute the correlation matrix using past trading volumes from the training data. This involves a total of 
$$
26 \times 10 \times N = 260N
$$ 
data samples, where \(N\) represents the number of stocks. We set the training process to be with 26 bins and 10 trading days \footnote{This is chosen through the autocorrelation analysis that has been detailed in \Cref{sec:data_featrues} }. After determining the correlation matrix, we employ Principal Component Analysis (PCA) to obtain an embedding, aiming to reduce the dimensionality and achieve an \(N \times N\) matrix. Based on the cumulative sum, which ranges between 0 and 1, we decide the number of eigenvectors $M$ to retain from the embedding, resulting in an \(N \times M\) matrix where \(M \leq N\). Following this, we decide on a method to categorize the stocks based on the dimensionality-reduced embedding matrix of shape \(N \times M\). Finally, we perform clustering using the K-means++ algorithm based on the M vectors from the embedding matrix.

\subsubsection{Universal Asset Model}

\begin{figure}[ht!]
    \centering
    \includegraphics[width=0.6\linewidth]{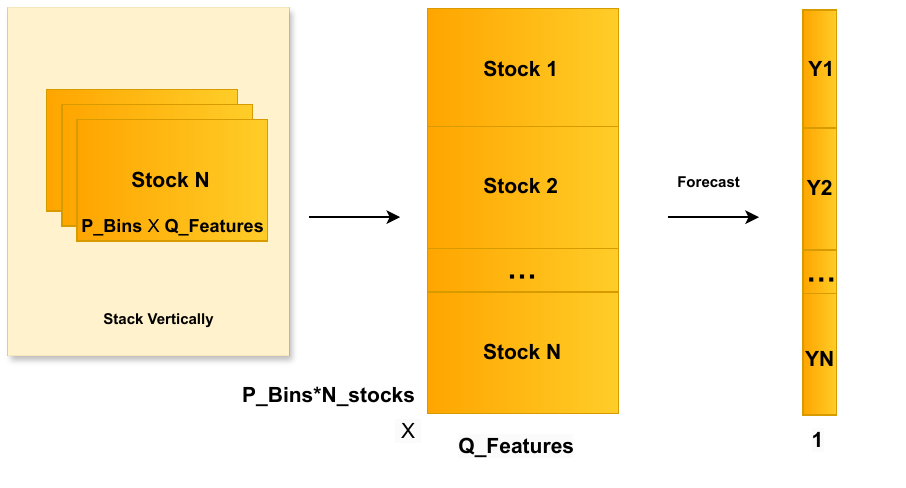}
    \caption{\textbf{Universal Model Diagram}. The left matrix is the predictors and the right matrix is the target, with shape to be $(N\_Bins*N\_stocks, N\_Features)$ and $(N\_Bins*N\_stocks, 1)$ respectively. } 
    \label{fig:universalDiagram}
    \end{figure}

Under the universal asset model scheme, we train a single model on the pooled dataset encompassing all stocks, leveraging each stock's historical features as predictors. The procedure of forecasting is illustrated in 
\Cref{fig:universalDiagram}. In this depiction, the input \(X\) to the regression models takes the form of a matrix with dimensions \((\text{NumBins} \times \text{NumStock}, \text{NumFeatures})\), while the output \(y\) is structured as \((\text{NumBins} \times \text{NumStock}, 1)\). This scheme allows the stacked matrix to share information across stocks, encapsulating both daily and intraday commonalities among them.

\subsection{Models}
This section provides an overview of the models utilized in our numerical experiments.

\subsubsection{Linear Regression}
\noindent \textbf{Ordinary Least Squares}

In the context of Ordinary Least Squares (OLS) regression, we apply the method to the set of predictors as follows. Let $\mathbf{u} = (u_1, u_2, \ldots, u_p)$ represent the vector of input predictors. The OLS model for predicting the volume during  $(t,t+h]$ for asset $i$ is given by:

\begin{equation}
    \text{Volume}_{i,t+h} = \alpha_i + \sum_{l=1}^{p} \beta_l u_l + \epsilon_{i,t+h}
    \label{eq:ols}
\end{equation}

\noindent \textbf{Least absolute shrinkage and selection operator}

When the number of predictors is close to the number of observations, or when there is a high correlation among predictor variables, the Ordinary Least Squares (OLS) model tends to overfit. This issue is especially pertinent to volume forecasting in our research, where predictors such as notional buy volume, notional sell volume, and traded amount may be correlated. 
To address this, the LASSO (Least Absolute Shrinkage and Selection Operator) regression method is employed, which mitigates overfitting by introducing a penalty on the coefficients of the regression model. LASSO serves two primary functions: it performs variable selection by shrinking some coefficients to zero, thereby excluding irrelevant predictors, and it regularizes the model to enhance both its predictive accuracy.

The LASSO objective function is a combination of the sum of squared residuals and an L1 penalty on the regression coefficients, as depicted in \Cref{eq:lasso}:

\begin{equation}
    L_{i} =  \sum_{t} \left[(\text{Volume}_{i,t+h} - \beta_0 - \sum_{j=1}^{p} \beta_j u_{j})^2  \right] + \lambda \sum_{j=1}^{p} ||\beta_j||_{1}
    \label{eq:lasso}
\end{equation}

Here, \( \lambda \) is a hyperparameter that controls the strength of the penalty. In our experiments, we explore a range of \( \lambda \) values and select the one that yields the best performance on the validation dataset for our forecasting model.

\medskip 

\noindent \textbf{Ridge Regression}

Ridge regression is a technique used to address the issue of multicollinearity in linear regression models. Multicollinearity occurs when predictor variables are highly correlated, leading to instability in the estimation of regression coefficients. Ridge regression mitigates this problem by adding a penalty term to the OLS objective function, which shrinks the regression coefficients toward zero.

The Ridge regression model for predicting the volume $\text{Volume}_{i,t+h}$ at time $t+h$ for asset $i$, using the vector of input predictors $\mathbf{u} = (u_1, u_2, \ldots, u_p)^T$, is given by:

\begin{equation}
    \text{Volume}_{i,t+h} = \alpha_i + \sum_{l=1}^{p} \beta_l u_l + \epsilon_{i,t+h}
    \label{eq:ridge}
\end{equation}

The objective function of Ridge regression is the sum of squared residuals with an added L2 penalty on the regression coefficients:

\begin{equation}
    L_{i} = \sum_{t} \left[(\text{Volume}_{i,t+h} - \alpha_i - \sum_{l=1}^{p} \beta_l u_l)^2 \right] + \lambda \sum_{l=1}^{p} \beta_l^2
    \label{eq:ridge_obj}
\end{equation}

Here, $\lambda$ is a hyperparameter that controls the strength of the penalty. A larger value of $\lambda$ imposes a greater penalty, leading to smaller coefficient values and reducing the model's complexity. In our experiments, we tune $\lambda$ to find the optimal balance between fitting the training data and maintaining model simplicity to prevent overfitting.

\subsubsection{Nonliner Models} \label{sec:CNN_LSTM}

\paragraph{XGBoost.} 

In addressing the limitations of linear models, it's crucial to consider non-linear models due to the potential non-linear relationships between the dependent variable and predictors, as well as interactions among predictors. One effective method to incorporate non-linearity and interactions is through decision trees, as detailed in the work of \cite{hastie2009elements}.  Among the most prominent decision-tree-based ensemble methods is XGBoost, an algorithm that harnesses the power of gradient boosting from \cite{chen2016xgboost}.

The objective function of XGBoost, which the algorithm aims to minimize, is given by:

\begin{equation}
    L(\Phi) = \sum_{i=1}^{n} l(\text{Volume}_i, \widehat{\text{Volume}}_i) + \sum_{k=1}^{K} \Omega(f_k)
    \label{eq:xgboost}
\end{equation}

In this equation, \(L(\Phi)\) represents the overall loss function, \(l(\text{Volume}_i, \hat{\text{Volume}}_i)\) denotes the loss function that measures the discrepancy between the predicted volume \(\hat{\text{Volume}}_i\) and the actual volume \(\text{Volume}_i\) for the \(i\)-th instance, and \(\Omega(f_k)\) is the regularization term that penalizes the complexity of the \(k\)-th tree \(f_k\) in the ensemble. This regularization term is crucial for mitigating overfitting, which is a typical challenge in decision tree models.

XGBoost optimizes this objective function by iteratively constructing a series of trees, each striving to rectify the errors of its predecessors. The regularization term serves to mitigate the overfitting of the model.

\begin{figure}
    \centering
    \includegraphics[width=0.85\linewidth]{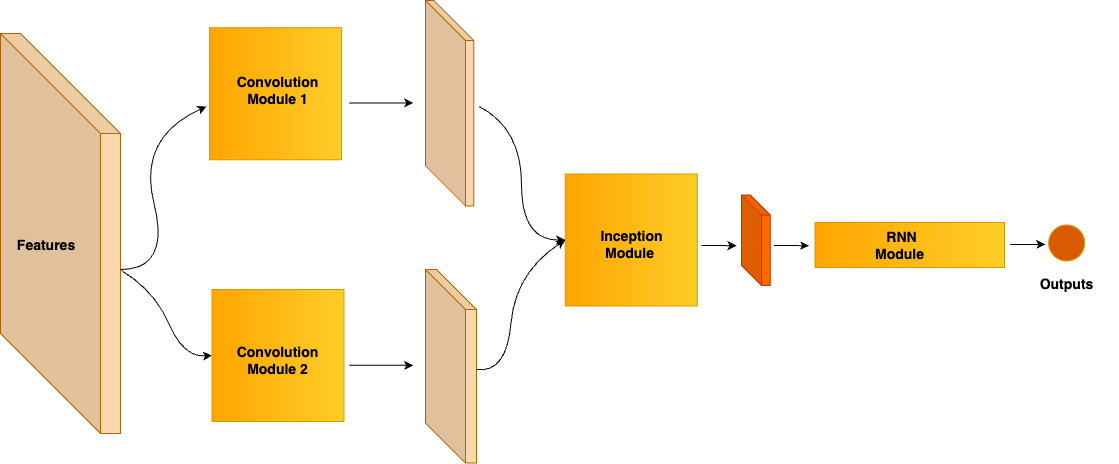}
    \caption{\textbf{Diagram of the  \textbf{$\text{DeepLOB}^{v}$} architechture}. The light yellow rectangles in the figure are the network module. The two Convolution modules are employed to extract information from the features. The Inception module is utilized to combine the information extracted. The LSTM module is then used to find the long-term dependencies and make predictions.}
    \label{fig:cnn_lstm}
\end{figure}

\paragraph{$\text{DeepLOB}^{v}$.} 
In this section, we employ Convolutional Neural Networks (CNN) for feature extraction and Long Short-Term Memory (LSTM) networks to address long-term dependencies within the dataset. In the Convolution module, we incorporate the Inception module, inspired by the DeepLOB architecture from \cite{deeplob}, which utilized one-dimensional CNN kernels. These kernels are applied both vertically and horizontally to capture temporal and spatial patterns, respectively. The outputs of the CNN module are then fed into an Inception module (\cite{szegedy_going_2014}) 
to combine the extracted latent vectors. \kl{The Inception module, as described by \cite{szegedy_going_2014}, allows for the efficient use of computational resources while increasing the depth and width of the network. It helps in capturing features at different scales simultaneously, which could be particularly useful for analyzing complex patterns in trading volume data.} Finally, a single-layer LSTM is utilized to process the correlations across different dates and deliver the forecast results. This integrated approach, combining the Convolution module, Inception module, and LSTM, is designed to effectively capture and analyze the complex dynamics of trading volume. We name this model as $\text{DeepLOB}^{v}$.\footnote{The superscript $v$ of the $\text{DeepLOB}^{v}$ represents the DeepLOB model with the network structure modified for the volume forecasting task.} \Cref{fig:cnn_lstm} shows the diagram of the network, illustrating this comprehensive structure.

\section{Experiments} \label{sec:methods}

\subsection{Model Performance} 

\label{sec:Overall_all_models }

\begin{table}[!ht]
    \centering
    \begin{threeparttable}
    \caption{\textbf{Overview of Model Performance Comparison}. The model with the best out-of-sample performance under different schemes is highlighted in \textcolor{red}{red}, while the best-performed model across all the schemes is marked in bold \textcolor{red}{\textbf{red}}. }
    \begin{tabular}{lllccc}
    \toprule
     &  &  & & \multicolumn{2}{c}{\({R}^2\)} \\
    \cmidrule(lr){5-6}
    \textbf{Predictors} & \textbf{Model} & \textbf{Features} & \textbf{Num}\tnote{a}& \textbf{Mean} & \textbf{Std} \\
    \midrule
    Benchmark & CMEM & CMEM Components & - & 0.265 & 0.105 \\
    \midrule
    SAM & RIDGE & CMEM Components\tnote{c} & 7 & 0.379 & 0.065\\
    & RIDGE & Auxiliary Predictors\tnote{b} & 47 & 0.405 & 0.063\\
    & RIDGE & Auxiliary Predictors \& CMEM Components\tnote{d}& 54 & 0.491 & 0.059 \\
    & XGB & Auxiliary Predictors \& CMEM Components & 54 & 0.532 & 0.081\\
    & $\text{DeepLOB}^{v}$& Auxiliary Predictors \& CMEM Components& 54 & \textcolor{red}{0.566} & 0.071 \\
    \midrule
    CAM & RIDGE & Auxiliary Predictors \& CMEM Components& 54 & 0.557 & 0.057 \\
    & XGB & Auxiliary Predictors \& CMEM Components& 54 & \textcolor{red}{0.600} & 0.083\\
    & $\text{DeepLOB}^{v}$ & Auxiliary Predictors \& CMEM Components& 54 & 0.595 & 0.071\\
    \midrule
    UAM & RIDGE & Auxiliary Predictors \& CMEM Components& 54 & 0.459 & 0.056\\
    & XGB & Auxiliary Predictors \& CMEM Components& 54 & 0.622 & 0.068 \\
    & $\text{DeepLOB}^{v}$ & Auxiliary Predictors \& CMEM Components& 54 & \textcolor{red}{\textbf{0.624}} & 0.073 \\

    \bottomrule
    \end{tabular}
    \begin{tablenotes}
    \item[a] Number of predictors.
    \item[b] For example, notional buy orders, notional sell orders, executed shares, etc. 
    \item[c] Daily component, intraday periodic component, intraday dynamic component. 
    \item[d] Here, we \mc{include one additional} feature for CMEM Components, which is produced by \mc{multiplying} the three components, namely $\eta \cdot s \cdot \mu, \eta \cdot s$, and $ s \cdot \mu, \eta \cdot \mu$. \mc{Altogether, there are seven} CMEM Components as predictors.
    \end{tablenotes}
    \label{tab:model_performance}
    \end{threeparttable}
    \end{table}

\begin{figure}
    \centering
    \includegraphics[width=1.0\linewidth]{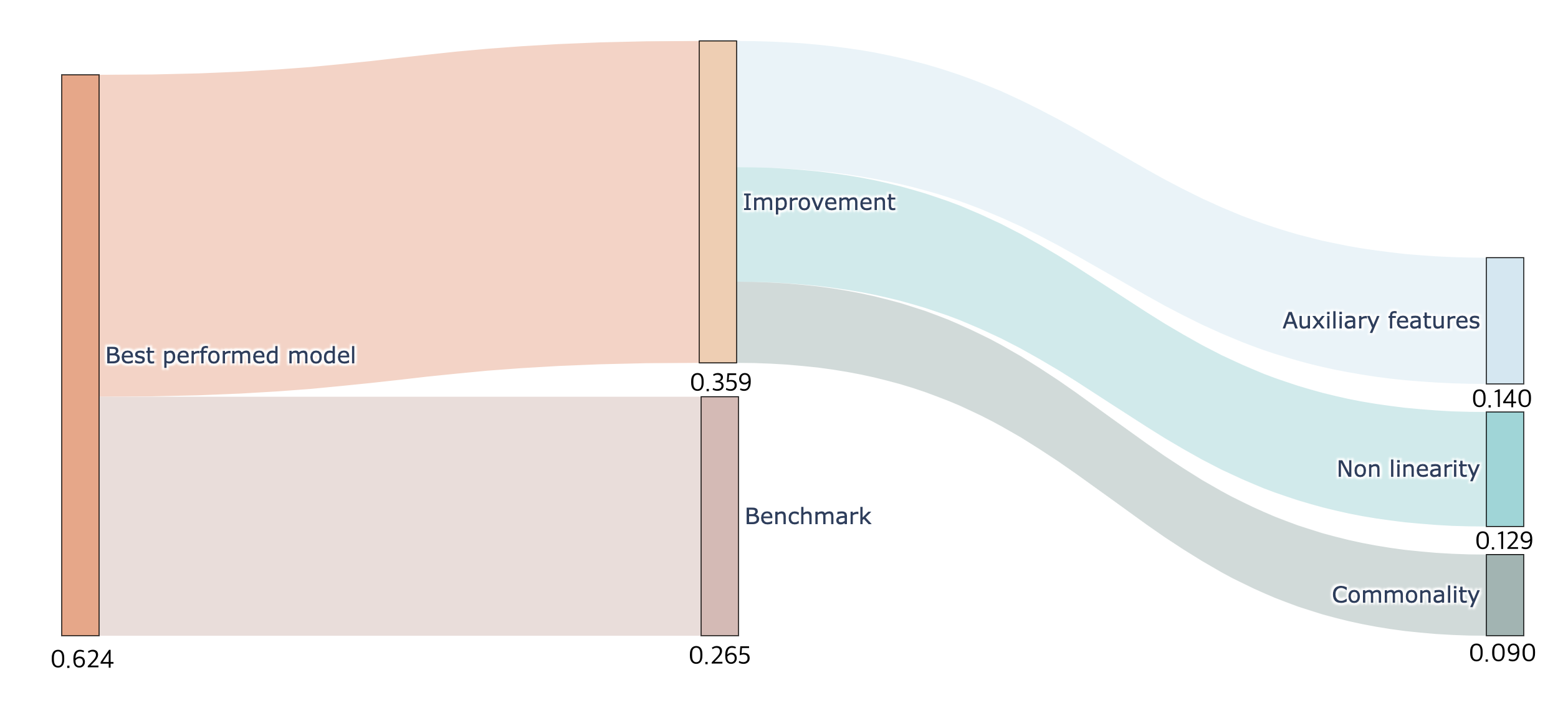}
    \caption{\textbf{A Sankey diagram illustrates the properties that contribute to improvements in prediction.} The marked \kl{numbers around the bars} are the out-of-sample $R^2$ of each model and the difference in $R^2$ between them.}
    \label{fig:sankey}
\end{figure}

As illustrated in \Cref{tab:model_performance}, we explore various models ranging from linear to nonlinear \kl{ approaches for intraday trading volume forecasting}. 
The baseline model is the CMEM, incorporating three key components: daily, intraday periodic, and intraday non-periodic, which are then multiplied, yielding an out-of-sample $R^2$ of \textbf{0.265}. By integrating Auxiliary predictors, uncovering intraday commonalities, and applying nonlinear models, we elevate this benchmark to an $R^2$ of \textbf{0.624}. We draw the following conclusions from \Cref{tab:model_performance}, \Cref{tab:models_r2} and \Cref{fig:sankey}.

We first begin with showing the benchmark performance.\footnote{The out-of-sample $R^2$ of the CMEM under the static prediction setting is 0.240. We observe an improvement when transitioning to the dynamic prediction setting, achieving an $R^2$ of 0.265. This result indicates that incorporating the newly updated observation for the day $t+1$ during prediction enhances the accuracy of volume forecasting. Given that experiments are most effective when conducted within the context of dynamic prediction, this approach will be employed for all models discussed in this paper.}  The result shows that the CMEM achieves an $R^2$ of 0.265. However, \cite{kf_cmem2016} mentioned that CMEM imposed positivity constraints on its three principal components and the noise term. This requirement further   
\kl{increases the complexity of parameter estimation} due to its inherent nonlinear structure, and \kl{furthermore} renders the predictions unstable. To avoid these issues, referring to the approach in \cite{kf_cmem2016}, we decompose the three components of the benchmark model and combine them \kl{linearly}. \kl{With these new features from the benchmark model,} \Cref{tab:models_r2} shows that the model performance is improved to 0.379 when Ridge is used as a regression model.

As previously stated in \Cref{sec:literature},  intraday volume exhibits a right-skewed distribution, as documented in \cite{volume_skewness}, with a long right tail and a few large observations. \kl{To address this issue, a logarithmic transformation is applied. It} compresses the range of values, reducing the impact of outliers, and thereby improving the prediction and enhancing stability. In our research, all other models take the logarithmic transformation of the predictors as the input predictors. \kl{Building on this preprocessing step, we use the RIDGE model with CMEM components as another benchmark. This allows us to evaluate the incremental predictive power gained from auxiliary predictors while accounting for the effects of the logarithmic transformation that we applied in our experiments. This comparison is important since the original CMEM only uses multiplicative components without the logarithmic transformation.}

\begin{table}[ht]
\centering
\caption{\textbf{$R^2$ values for different models with auxiliary predictors.} The model with the best out-of-sample performance is highlighted in bold font.}
\begin{tabular}{lccc}
\toprule
 & & \multicolumn{2}{c}{\( R^2 \)} \\
\cmidrule(lr){3-4}
Model & Predictors & Mean & Std \\
\midrule
OLS &  CMEM Components & 0.277 & 0.071 \\
LASSO & CMEM Components & 0.379& 0.065 \\
RIDGE & CMEM Components & \textbf{0.379} & 0.065 \\
\midrule
OLS & Auxiliary Predictors & 0.400 & 0.062 \\
LASSO & Auxiliary Predictors & 0.403 & 0.063 \\
RIDGE & Auxiliary Predictors & \textbf{0.405} & 0.063 \\
\midrule
OLS & Auxiliary Predictors \& CMEM Components& 0.484 & 0.061 \\
LASSO & Auxiliary Predictors \& CMEM Components& 0.487 & 0.059 \\
RIDGE & Auxiliary Predictors \& CMEM Components& \textbf{0.491} & 0.059 \\
\bottomrule
\end{tabular}
\label{tab:models_r2}
\end{table}

Regarding auxiliary predictors and linear models, we observe that, on one hand, the prediction with auxiliary predictors achieved an $R^2$ of 0.405, which outperformed the prediction with solely CMEM components, which achieved 0.379. This suggests that the auxiliary predictors were highly effective. We speculate that the auxiliary predictors enhance prediction stability by offering extra information, thereby expanding the scope beyond mere univariate time series forecasting. On the other hand, when the predictors used in the prediction are CMEM and the auxiliary predictors, we found that the $R^2$  is improved to 0.491, and the standard deviation is reduced from 0.065 to 0.059. This demonstrates that the addition of auxiliary predictors to the prediction not only significantly enhances the accuracy of the prediction but also renders the prediction more stable. \kl{A plausible hypothesis for this observed phenomenon is that when the decomposition predictors derived from CMEM exhibit suboptimal performance, thereby negatively impacting the overall predictive accuracy and stability, the auxiliary predictors compensate for these deficiencies, consequently becoming the primary contributors to successful predictions.}  
In addition, we also observe a higher out-of-sample $R^2$ in LASSO and Ridge compared to OLS, suggesting that regularization does further aid performance.

From the above analysis, we conclude that \textbf{auxiliary predictors} play a critical role in enhancing our model's predictive capability. These predictors provide supplementary information, such as the trading amount from both buying and selling sides over various periods ranging from half an hour to two hours, and the number of orders during specific intervals of the day—open, mid-day, and day-close intervals. On the other hand, these auxiliary predictors aid in correcting our model when the CMEM components inaccurately assess the situation. Upon examining numerous instances where the CMEM-based predictions perform poorly within a day, we observe that the trading volume in these poorly-\mc{performing} samples did not always conform to the anticipated U-shape, particularly during the close interval, which are flattened curves. Due to CMEM's inherent design to depict a U-shape, it leads to significant discrepancies between predicted and actual values at such data points. Consequently, by addressing these discrepancies, our model not only witnesses an improvement in the mean of $R^2$ but also experiences a reduction in the variance of the out-of-sample $R^2$ across dates. We successfully mitigated the issue of outliers\footnote{Directly using CMEM Components in conjunction with linear regression resulted in some days' predictions having an $R^2$ even below -1.
},
a problem that could arise when exclusively relying on CMEM components for predictions.

Generally speaking, there are \mc{significant} improvements when moving from linear models to nonlinear models, over SAM, CAM, and UAM training schemes. For instance, in UAM, the $R^2$ of the linear model is 0.549, while the $R^2$ of the nonlinear model is 0.622.   It is worth noting that the $\text{DeepLOB}^{v}$ network structure we designed performed best in the SAM setting, achieving an $R^2$ of 0.566, which is higher than the best-performing linear model of 0.491. It is also the state-of-the-art model in SAM, which shows that the neural network structure we designed is very effective for the task of predicting the intraday volume.

The superior performance of nonlinear models can be attributed to the inherent nonlinearity of trading activity. Market dynamics are often characterized by intricate patterns and complex interactions between various factors \mc{and market conditions}, leading to nonlinear relationships between trading volume and its predictors. Several phenomena, such as the herding effect \citep{herd}, overreaction and reversal  \citep{darratIntradayVolumeVolatility2007}, and the spillover effect  \citep{zhang_forecasting_2025}, exemplify the nonlinear behavior of market participants.\footnote{The herding effect describes the tendency of investors to mimic and follow the crowd, resulting in clustered and amplified trading activity. For instance, when large institutional investors or renowned traders enter a position, it can trigger a wave of retail investors following suit, leading to a surge in intraday trading volume within a short period. Conversely, if institutional players start selling, it may instigate a frenzy of panic selling, causing a spike in daily traded volumes.}

Moreover, overreaction to new information can cause short-term volume volatility, followed by subsequent reversals as the market corrects itself.\footnote{For example, when a company releases positive news, it may immediately attract a flurry of buy orders, resulting in a substantial increase in intraday trading volume. However, as investors rationally analyze the information, trading activity may revert to normal levels, leading to a decline in volume.} Additionally, negative news can rapidly propagate through the market, triggering widespread panic selling, known as the spillover effect, which has been thoroughly documented by \cite{zhang2023graph}. A significant adverse event could lead to a market-wide sell-off, resulting in a significant surge in daily trading volumes. The emergence of these nonlinear behavioral patterns suggests that the factors influencing trading volume exhibit complex dynamic relationships, which linear models struggle to capture effectively. By uncovering the hidden nonlinear patterns in the data, machine learning algorithms can potentially forecast intraday trading volume fluctuations more accurately, as demonstrated by the results of our numerical experiments.

\begin{figure}[ht]
    \centering
    \begin{subfigure}[b]{0.52\linewidth}
        \centering
        \includegraphics[width=\linewidth]{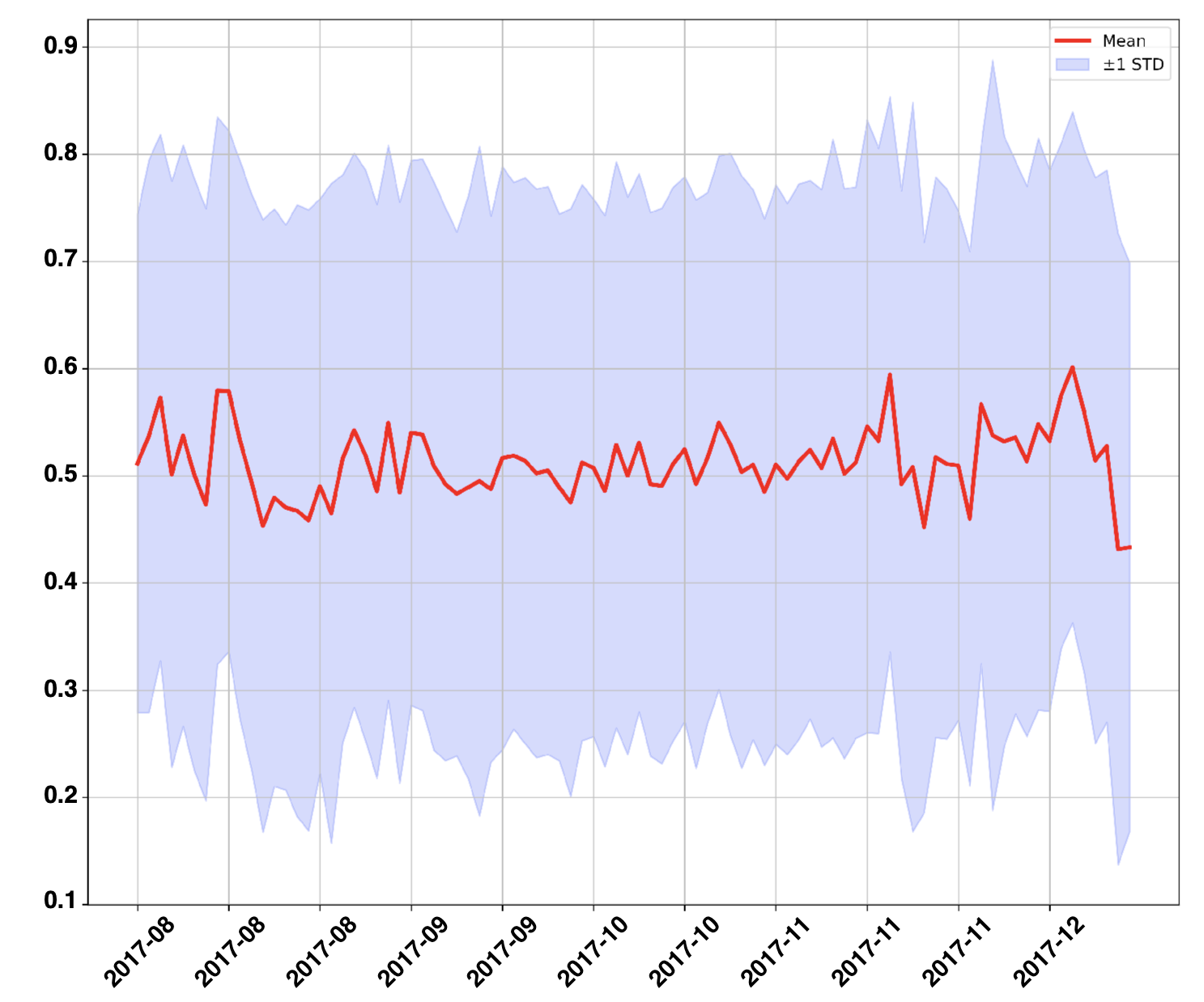}
        \caption{\textbf{Sensitivity Analysis across Dates}. }
        \label{fig:date_stability}
    \end{subfigure}
    \hfill
    \begin{subfigure}[b]{0.47\linewidth}
        \centering
        \includegraphics[width=\linewidth]{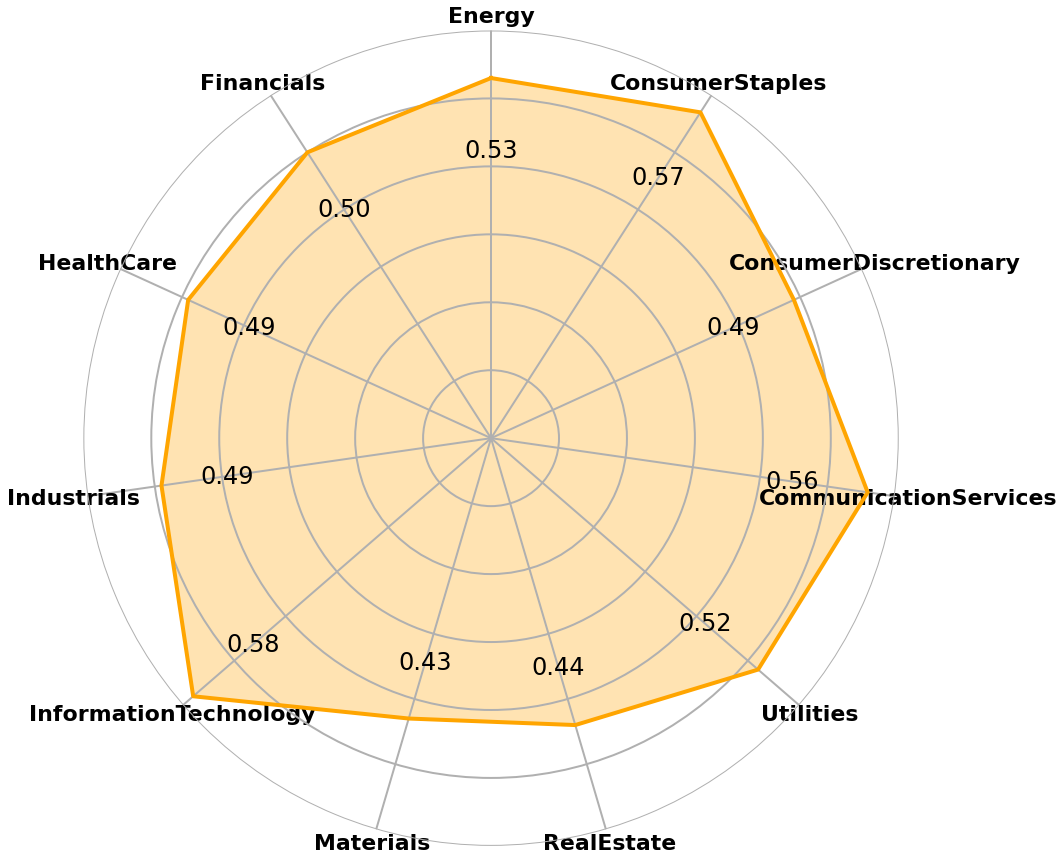}
        \caption{\textbf{Sensitivity Analysis across Sectors}}
        \label{fig:prediction_stability}
    \end{subfigure}
    \caption{Comparing stability across different conditions. In the left figure, the solid blue line is the mean of out-of-sample $R^2$, and the shadow area is one standard deviation. In the right figure, the solid blue line is the mean of out-of-sample $R^2$, and the shadow area is one standard deviation.}
    \label{fig:combined_stability}
\end{figure}

An interesting question to investigate is whether the commonality between stocks improves the prediction. To this end, we present the results of CAM and UAM in \Cref{tab:model_performance}. For the XGB model, the transition from a SAM scheme to a UAM scheme results in an improvement in $R^2$ from 0.532 to 0.622. This gain highlights the presence of commonalities in trading volumes across stocks and suggests that integrating data from multiple assets can reveal underlying patterns not apparent when analyzing assets independently. To further explore the concept of commonality, we propose a more nuanced approach.  Rather than aggregating all stocks into a single group, an alternative approach is to categorize stocks based on their proximity and similarity, which \kl{is in line with} 
the CAM setting. Interestingly, for the linear model, the results demonstrated that CAM outperformed UAM (with respective values of 0.577 and 0.549). This corroborates our previous hypothesis, suggesting that a more refined grouping methodology can facilitate more accurate commonality discovery, thereby enhancing prediction accuracy. Further analysis of the CAM is provided in \Cref{sec:cluster_asset_model}.

Potential reasons for the improvement observed when comparing the SAM to CAM or UAM schemes include hedging strategies involving stocks, which can entail positions across different stocks. For instance, if investors are concerned about a potential decline in a particular industry, they might opt to sell stocks within that industry while simultaneously buying stocks from another industry.  Another example, which commonly results in the concurrent purchase of other stocks, is a strategy that involves purchasing stocks within a specific time frame to hedge out beta and capture alpha. These all lead to a commonality in intraday trading volume, reflecting the market structure, which can in turn facilitate the prediction of trading volumes.

In addition, we are concerned with the stability of our predictions, which we examine from two perspectives. On one hand, we are interested in the stability over time. \Cref{fig:date_stability}  shows the out-of-sample $R^2$ on the test dataset. We observe that the performance is very stable overall. It is noteworthy that there were some fluctuations at the end of the year. Our speculation and explanation is that these fluctuations may be caused by large fluctuations in liquidity and trader behavior around Christmas.  On the other hand, we are concerned with the predictive ability of different stock sectors. \Cref{fig:prediction_stability} shows the differences in the predictive ability of our model between different sectors. Most of the differences are not large, except for some specific industries such as Materials and Real Estate.

\subsection{Feature Importance Analysis}

In this section, we discuss the feature importance across different models and identify the top predictors contributing to the forecasting task. In the context of OLS regression, the $t$-value is a crucial statistic for assessing the importance of individual predictors. \kl{It measures how many standard deviations an estimated coefficient is from zero, calculated as the ratio of the coefficient to its standard error.} \kl{Higher absolute $t$-values indicate statistically significant features that substantially explain variability in the dependent variable.} \kl{In the context of LASSO regression, the magnitude of coefficients directly indicates feature importance. Note that all features have been standardized prior to entering the regression model. Larger absolute coefficient values suggest a stronger influence on the target variable, with the regularization process automatically shrinking less important features toward zero.}

\begin{table}[ht]
    \centering
    \caption{\textbf{Summary of the top 5 most important predictors for each model considered}; $x$(\texttt{x}) is the multiplication of $\eta$ (\texttt{eta}), $s$ (\texttt{seas}) and $\mu$ (\texttt{mu}) from \eqref{eq:cmem}, where, \texttt{eta} is the daily component of volume,  \texttt{seas} is the intraday periodic component of volume, and \texttt{mu} is the intraday non-periodic component of volume. The variable \texttt{ntn} denotes the trading amount. For additional details on the predictors, please refer to Table \ref{tab:feature_importance_explain}.}
    \begin{tabular}{lllllll}
    \toprule
     & \textbf{OLS} & \textbf{LASSO} & \textbf{Ridge} & \textbf{XGB} & \textbf{NN} \\
     & t-value & coefficient & coefficient & Weights & SHAP \\
     \midrule
    1 & \texttt{seas} & \texttt{x} & \texttt{intraday\_ntn} & \texttt{x} & \texttt{x} \\
    2 & \texttt{x} & \texttt{seas}

    & \texttt{x} & \texttt{eta$\cdot$seas} & \texttt{eta$\cdot$seas} \\
    3 & \texttt{volSellQty} & \texttt{daily\_volBuyQty} & \texttt{daily\_qty} & \texttt{mu} & \texttt{ntn\_8} \\
    4 & \texttt{volBuyNotional} & \texttt{daily\_volSellQty} & \texttt{seas} & \texttt{nrTrades} & \texttt{seas} \\
    5 & \texttt{intraday\_qty} & \texttt{volBuyNotional\_2} & \texttt{volBuyQty} & \texttt{eta} & \texttt{ntn} \\
    \bottomrule
    \end{tabular}
    \label{tab:feature_importance}
    \end{table}

\kl{Moving from the linear models to nonlinear approaches, we apply both tree models and neural networks in our analysis. In the context of tree models,} we utilize the XGBoost framework to assess the significance of predictors within our model. Specifically, we \kl{utilize} the "weight" metric, which quantifies the frequency with which a feature is used to split the data across all decision trees in the ensemble, as the criterion for determining feature importance. \kl{For neural network} models, particularly those that integrate CNN and LSTM, Shapley Additive explanations (SHAP) is a popular method for interpreting the outputs of the model. \kl{Derived from cooperative game theory, SHAP values provide a framework for fairly attributing contributions among features. The absolute magnitude of these values effectively indicates each predictor's importance in the forecasting process.} 

From 
\Cref{tab:feature_importance}, we observe that the decomposition components consistently rank as the most influential, with several auxiliary predictors also appearing among the top five, highlighting the effectiveness of our proposed predictors in the model. \mc{A description} of the predictors in \Cref{tab:feature_importance} is provided in \Cref{tab:feature_importance_explain}. 
Moreover, hourly, intraday, and daily aggregated predictors demonstrate notable efficacy, indicating that capturing temporal dependencies at various scales is crucial, since trading volume patterns are influenced not only by immediate past conditions but also by broader temporal trends.

\section{Application of Forecasting Intraday Volume to VWAP Strategies} \label{sec:application}

\mc{A key use of intraday volume forecasts is in implementing VWAP replication strategies, as discussed} in \cite{dynamicVolume} and \cite{Brownlees_CMEM}, which aims to execute a large order over a specific period while minimizing market impact and achieving an average price that closely matches the VWAP for that period. A typical approach to replicate VWAP is to slice a large order into smaller pieces and execute them at specific time intervals throughout the trading period, with the size of each slice proportional to the \kl{expected trading} volume during the interval. This approach is proven in \cite{Brownlees_CMEM} to be able to reduce the price impact that \kl{would otherwise result from executing a single large order.}

\begin{table}[!htbp]
\centering
\caption{\textbf{Sampled Stocks from Stocks Pool Sorted by Trading Volume} We sort the stocks by trading volume, divide them into five groups, and randomly sample stocks from each group.}
\label{tab:stock_symbols}
\begin{tabular}{lll}
\toprule
\textbf{Exchange} & \textbf{Ticker} & \textbf{Name} \\
\midrule
NYSE & MGM & MGM Resorts International \\
NASDAQ & FITB & Fifth Third Bancorp \\
NYSE & GS & Goldman Sachs Group Inc. \\
NASDAQ & AEP & American Electric Power Company Inc. \\
NYSE & HP & Helmerich \& Payne Inc. \\
\bottomrule
\end{tabular}
\label{tab:names}
\end{table}

\Cref{tab:names} shows the stocks that are tested in this section. To evaluate the effectiveness of the proposed intraday volume forecasting models in practical trading applications, we analyze their performance in minimizing the VWAP order execution risk. The VWAP tracking error, measured in \mc{basis points} \kl{(bp)}, is employed as the error metric. We employ the definition of tracking error stated in \cite{kf_cmem2016} as follows

\begin{equation}
w_{t, i}=\frac{\text { volume }_{t, i}}{\sum_{i=1}^I \text { volume }_{t, i}},
\end{equation}

\begin{equation}
\begin{aligned}
\mathrm{VWAP}_t & =\frac{\sum_{i=1}^I \text { volume }_{t, i} \times \text { price }_{t, i}}{\sum_{i=1}^I \text { volume }_{t, i}} \\
& =\sum_{i=1}^I w_{t, i} \times \text { price }_{t, i},
\end{aligned}
\label{eq:vwap}
\end{equation}

\begin{equation}
\mathrm{VWAP}^{\mathrm{TE}}=\frac{1}{D} \sum_{t=1}^D \frac{\mid \mathrm{VWAP}_t-\text { replicated } \mathrm{VWAP}_t \mid}{\mathrm{VWAP}_t}.
\label{eq:TrackingError}
\end{equation}

\kl{In \Cref{eq:vwap}}, ${\text{price}_{t, i}}$ denotes the last recorded transaction price within interval $i$ on day $t$, serving as a proxy for the VWAP of that interval. The term $w_{t, i}$ represents the volume weight assigned to this interval. Each trading day is divided into 15-minute bins, therefore, $I = 26$ according to the NASDAQ and NYSE trading timetable. Please note that our model used here is the best-performed model proposed in \Cref{tab:model_performance}, i.e. $\text{DeepLOB}^{v}$ model with auxiliary predictors and CMEM components as input predictors and trained under the UAM scheme with dynamic prediction. The details for conducting dynamic prediction in this testing experiment \mc{are deferred to}  Appendix \ref{sec:application_details}.

\begin{figure}[!htbp]
    \centering
    \includegraphics[width=0.75\linewidth]{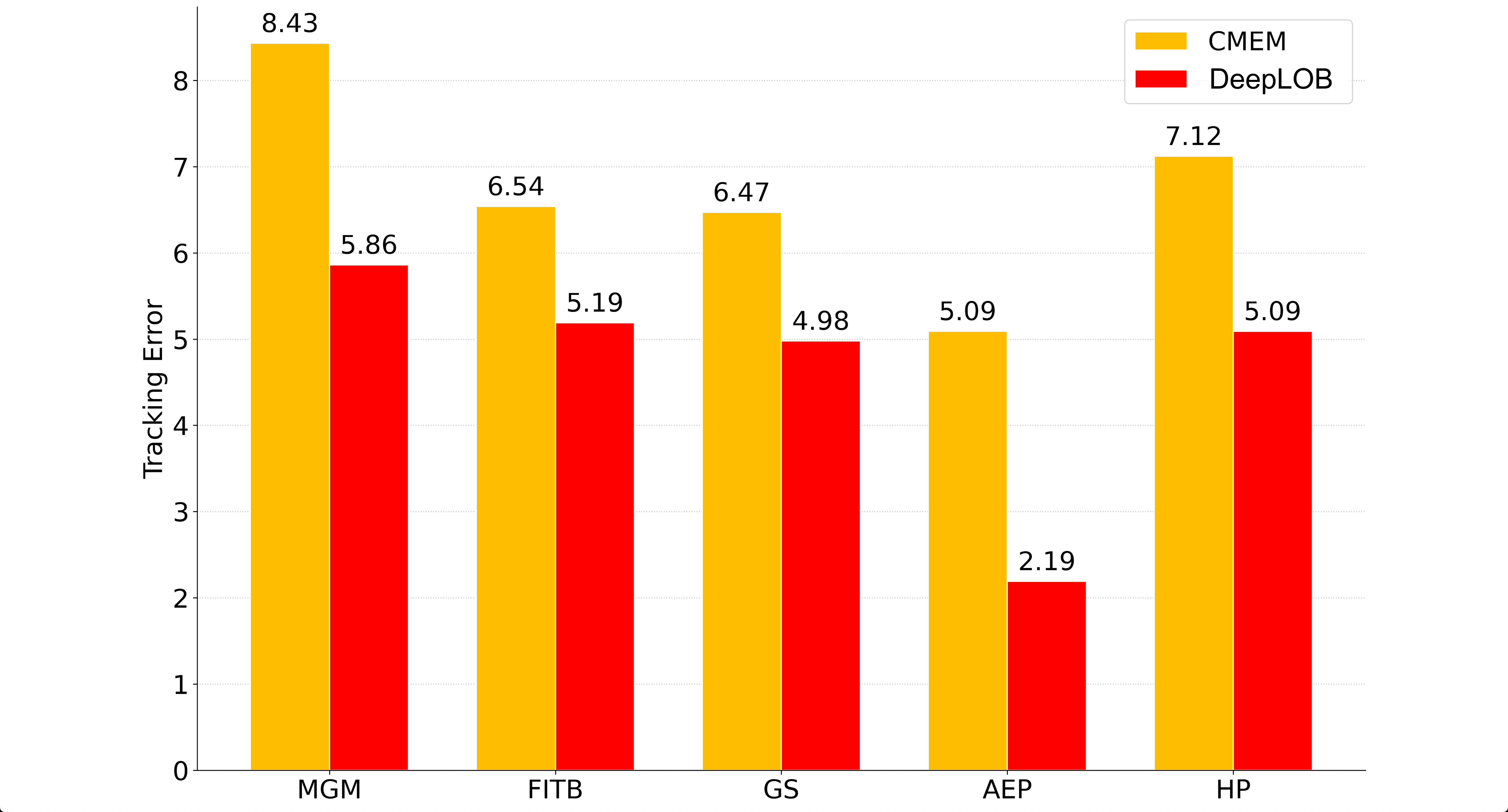}
    \caption{\textbf{Tracking Error Comparison}. 
    \kl{Comparison of tracking errors between the CMEM and our UAM $\text{DeepLOB}^{v}$ model across different tickers. The tracking error is measured in basis points (bp), where 1 bp equals to 0.01\%.}}
    \label{fig:te}
\end{figure}

As illustrated in \Cref{fig:te}, our model achieves performance improvements in tracking error across all five sampled stocks when compared with the CMEM benchmark. The most significant improvement is observed in AEP, with tracking errors decreasing by 2.90 \mc{basis} points, equating to enhancements of 57.0\%. Overall, the aggregate improvement of our model over the benchmark amounts to a notable average of \textbf{28.7\%}.\\

\kl{With} a more accurate forecasted intraday volume, its application in VWAP replication strategies can not only reduce tracking error but also increase the fill ratio of passive orders. There are many instances where improving the fill ratio of passive orders is \mc{beneficial}. For example, transitioning from the position held on the previous day to the anticipated position presents an optimal scheduling challenge. This process involves breaking down large orders into smaller ones, which may sometimes be passive orders. \cite{abhyankarBidaskSpreadsTrading1997} indicated that high trading volume was often associated with high liquidity. Meanwhile, placing passive orders when liquidity is high increases the likelihood of execution and helps achieve more favorable execution prices, as noted by \cite{percentageVolumeSVM}. As also mentioned in \Cref{sec:intro}, higher liquidity is associated with improved fill rates. By forecasting the intraday trading volume curve, one can submit more passive orders during periods of high volume and fewer during periods of low volume, to align with the strategy of placing more orders during peaks in the volume curve. We conducted the following experiments to test this hypothesis.

In our following experiments, we place passive limit orders\footnote{We choose to place the best passive orders, whose prices correspond to the best level ask or bid prices.}, and the size of the orders in each bin is in alignment with the volume curve. Specifically, for a given trading task, such as trading 1\% of the day's volume or a specified number of shares, the parent order with the total number of shares to be traded is then divided into smaller child orders for each time bin. The number of shares for each bin is determined by the proportion of the forecasted intraday volume to the total volume of the day. A matching engine is employed to generate execution results for subsequent analysis of executed orders. We calculate the fill ratio based on the recorded trades, which is determined in the following way

\begin{equation}
\mathrm{FillRatio}=\frac{\sum_{i}^{26}\sum_{t}^{T_i}{ \mathrm{ExcutedQuant}_{i,t}}}{\text{Quantity\_to\_Execute}},
\label{eq:TotalRevenue}
\end{equation}

\noindent 
where $i$ \mc{indexes} the bins, and $t$ the step inside each bin. There are 26 bins each day, with each bin containing a different number of steps, which are denoted as $T_{i}$\footnote{$T_{i}$ is determined by the amount of orders in the historical data. Our step definition encompasses one order from us at the step initiation, followed by 100 orders from other market participants. Under this setting, for a given 15-minute bin, the total number of steps $T_{i}$ is a variable.} 
The remaining quantity during each bin would be set as market orders at the end of each bin, thereby ensuring the complete execution of the parent order.

\begin{figure}[!ht]
    \centering
    \includegraphics[width=0.75\linewidth]{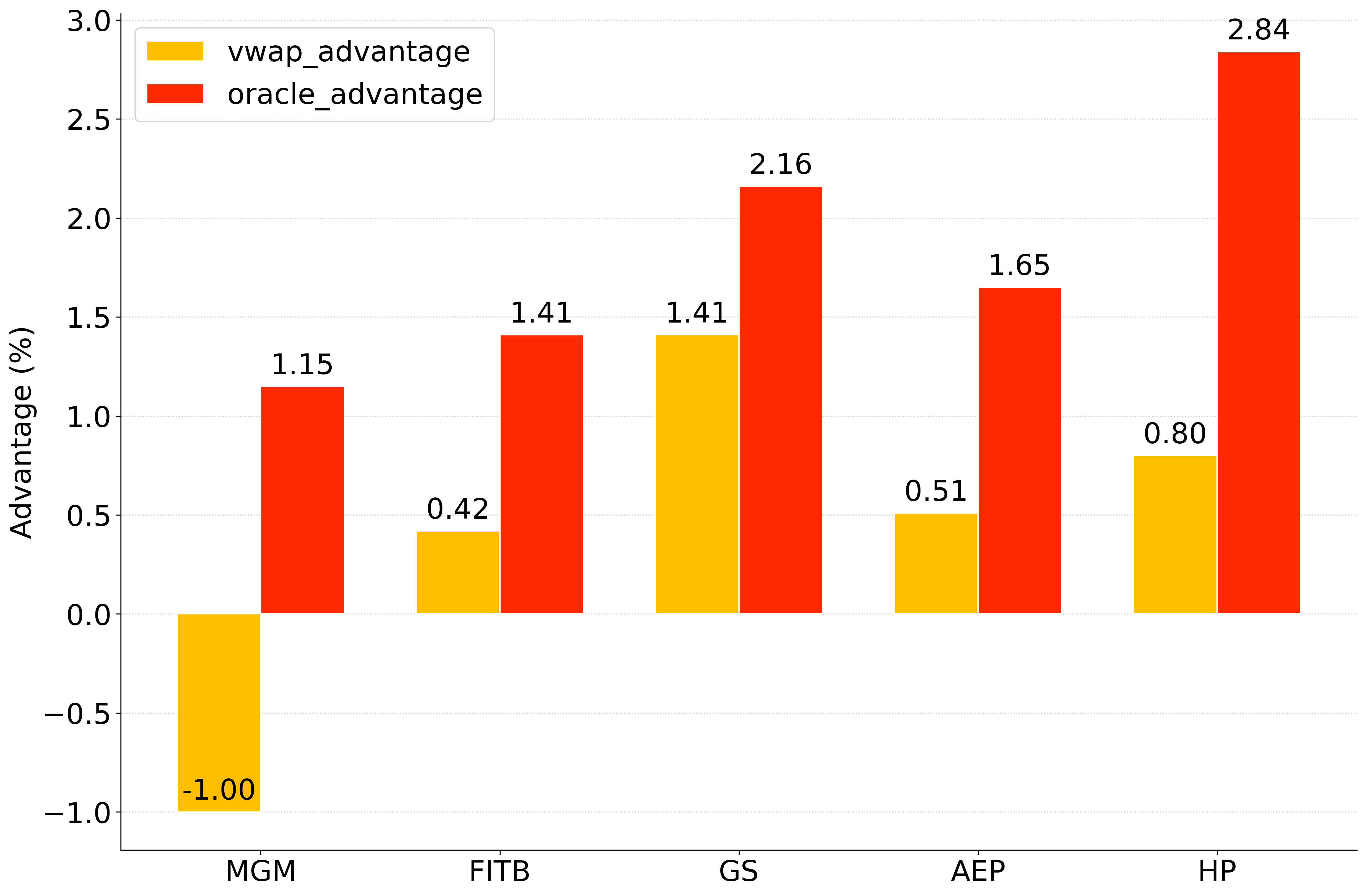}
    \caption{\textbf{Comparison of the Relative Fill Ratio}. The yellow bars \mc{denote} the advantage of our proposed model over the CMEM model; the orange bars denote the advantage of the oracle model over the CMEM model.}
    \label{fig:fulfillment_ratio}
\end{figure}

Subsequently, we compare the fill ratio using different order splitting policies: volume splitting based on the forecasted intraday volume from \textbf{CMEM}, the forecasted intraday volume from \kl{our \textbf{UAM DeepLOB$^{v}$} model}, and the forecasted intraday volume from the \textbf{oracle} volume. Afterward, we assess the advantages of the latter three models over the RM and compare their relative performance, as shown in \eqref{eq}, where $m$ denotes the name of the model to be compared, including \mc{our model UAM DeepLOB$^{v}$, and the oracle model}

\begin{equation}
    \text{Advantage} = \frac{\text{FillRatio}_{m} - \text{FillRatio}_{\text{CMEM}}}{\text{FillRatio}_{\text{CMEM}}}. 
\label{eq}
\end{equation}

As shown in  \Cref{fig:fulfillment_ratio}, our  
\kl{UAM DeepLOB$^{v}$} model demonstrates a fill ratio advantage over the RM model that is close to the advantage exhibited by the oracle model. Our model outperforms the RM model with an average advantage of 33.5 bp in achieving a higher fill ratio. The oracle model, on the other hand, has a 159.3 bp advantage over the CMEM model. This indicates that our model effectively places passive orders with a higher fill ratio, particularly when market liquidity is higher. This is achieved by submitting more orders during periods of high trading volume, thus leaving fewer shares to be executed during low-volume periods.

\section{Conclusion} \label{sec:conclusion}

In this study, we pioneer the identification of commonalities within intraday volume forecasting, employing clustered models to delve deeper into the intricate relationships between volume and auxiliary features. This approach marks a promising advancement in the field, offering new insights into the dynamics of intraday trading volume. We also introduce an innovative application of Deep Neural Network models, specifically employing a DeepLOB-style framework, to uncover the nonlinear characteristics inherent in volume forecasting. To the best of our knowledge, this research also presents the first framework utilizing a matching engine to examine \mc{and quantify} the influence of liquidity in the time-space continuum on execution outcomes, \mc{and to demonstrate its economic benefits}. This approach enables tracking the fill ratio of orders, as opposed to relying on best-level prices or the last traded price as the proxy of the executed prices, which may not accurately reflect executed transactions.

In future research, the intraday volatility could also be \kl{incorporated} to \kl{enhance} volume forecasting, as they were shown to be correlated in \cite{darrat2003intraday}, which documented a strong positive correlation between volume and volatility, with both exhibiting distinct U-shaped intraday patterns. More informative auxiliary features can also be beneficial to facilitate better forecasting. In addition to time-based volume, the volume profile, which represents volume in price space \kl{rather than time space}, is also crucial for intraday trading, \kl{where time space captures volume at different time intervals and price space represents volume across different price levels}. This provides auxiliary information by indicating price levels with high activity and offering insight into price discovery.

\newpage
\bibliographystyle{ACM-Reference-Format}
\bibliography{main}

\newpage
\appendices
\newpage
\newgeometry{margin=1.2in}
\section{Experimental Details of Forecasting Models}
\label{sec:experimental_details}

\subsection{Description of CMEM forecasting} \label{sec:apendix_details_of_cmem}
To improve the stability and robustness of predictions against outliers, one needs to mitigate the impact of data points that fall outside the typical range. Specifically, we define outliers as those observations that exceed or fall below the range determined by the maximum and minimum volume values observed in the input data from day \(t-D\) to day \(t\). Outliers will be clipped to the nearest boundary, either the maximum or minimum value observed in the data range. In addition, we apply minmax normalization to the input data. Minmax normalization is particularly common in financial datasets, as it preserves the relative relationships between values while scaling them to a consistent range, which is crucial for comparability and stability in financial analyses. This normalization is crucial for preparing data for model inputs. For example, in linear regression, it helps maintain numerical stability and ensures that each feature contributes proportionally to the prediction, avoiding bias towards features with larger scales. Furthermore, by normalizing the data, the coefficients of variables in LASSO regression can directly indicate the importance of each feature, facilitating the interpretation of feature importance. 

The input to the CMEM model is the intraday trading volume, split into 26 bins,  
with a series length of $D=10$ days, corresponding to 260 bins, while the output is the volume for the next 26 bins. The CMEM is a univariate time-series model. We normalize the volume by the outstanding shares, in line with  \cite{dynamicVolume}, resulting in a variable within the CMEM represented as $\frac{\text{Volume}_{i,t}}{\text{Outstanding\_Shares}_t}$, where $t$ indicates the day and $i$ denotes the bin. After the prediction step, the output is converted back to the number of shares. The CMEM's decomposition, along with auxiliary features such as trading amount, number of executed shares, and number of orders, is subsequently used as input features to various models for making predictions, ranging from linear regression to neural networks.

\subsection{Static and dynamic prediction} \label{ProblemSetup}

\begin{figure}[!htbp]
    \centering
    \includegraphics[width=0.65\linewidth]{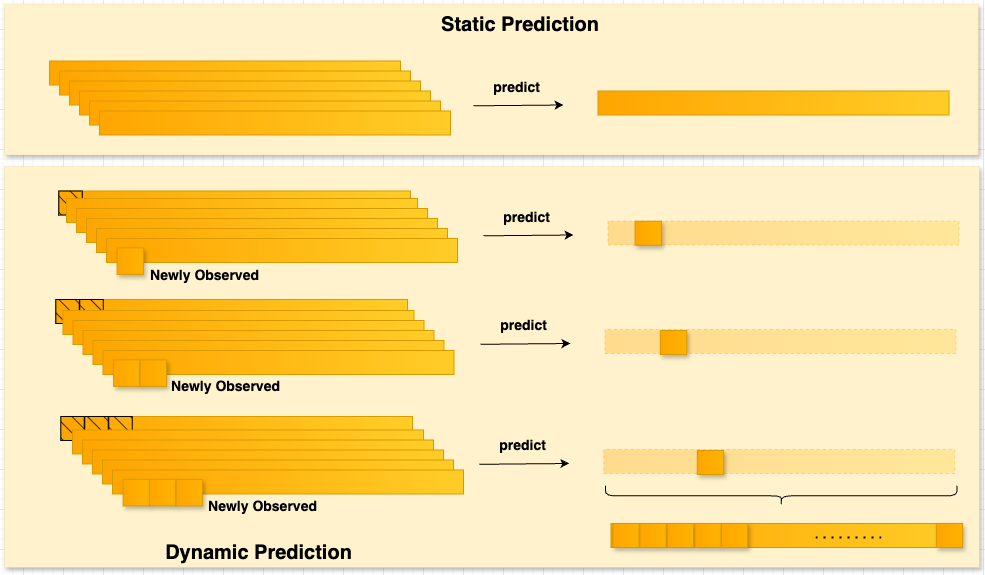}
    \caption{\textbf{Static and Dynamic Prediction}. The upper diagram illustrates static prediction, where the left side represents the input from the past several days, specifically from day $t$ to day $t-k$, and the right side shows the predicted volume for the next day, namely day $t+1$. The lower diagram depicts dynamic prediction, where prediction is repeated 26 times in a rolling window manner. The input incorporates newly observed predictors, and each time, the output is only one bin. The predicted 26 bins are then concatenated to form one output, which is the predicted intraday volume of the next day, namely day $t+1$.}
    \label{fig:static_dynamic}
\end{figure}

In our numerical experiments, we examine two settings for intraday volume forecasting, namely the static and dynamic prediction settings, as defined by \cite{dynamicVolume} and \cite{kf_cmem2016}. Static prediction involves forecasting the volume for all \(I\) bins of day \(t+1\) utilizing only the information available up to day \(t\). On the other hand, dynamic prediction refers to a one-bin-ahead forecasting approach, where the volume of a specific bin is predicted based on all available information up to the immediately preceding bin. \Cref{fig:static_dynamic} illustrates the procedure of these two prediction settings. 

The static VWAP replication strategy assumes
that the order slicing is set before the market opening and is not revised during the day. We adopt the definition in \cite{kf_cmem2016}, where the static and dynamic weights are implemented as
\begin{equation}
\hat{w}_{t, i}^{(s)}=\frac{\widehat{\operatorname{volume}}_{t, i}^{(s)}}{\sum_{i=1}^I \widehat{\operatorname{volume}}_{t, i}^{(s)}}, 
\end{equation}

\noindent where $\widehat{\operatorname{volum}}_{t, i}{ }^{(s)}$ denotes the volume from static predictions. In the dynamic VWAP replication strategy scenario, order slicing is updated at the frequency of each bin when a new intraday volume is progressively observed. The dynamic VWAP replication strategy employs the weights as defined below during slicing.

\begin{equation}
\hat{w}_{t, i}^{(d)}= \begin{cases}\frac{\widehat{\operatorname{volume}}_{t, i}^{(d)}}{\sum_{j=i}^I \widehat{\operatorname{volume}}_{t, i}^{(d)}}\left(1-\sum_{j=1}^{i-1} \hat{w}_{t, j}^{(d)}\right), & i=1, \ldots, I-1 \\ \left(1-\sum_{j=1}^{I-1} \hat{w}_{t, j}^{(d)}\right), & i=I,
\end{cases}
\end{equation}

\noindent where $\widehat{\operatorname{volum}}_{t, i}{ }^{(d)}$ is the volume from dynamic predictions. This weight is calculated as the ratio of the predicted volume in bin $i$ to the total predicted volume for the remaining bins, multiplied by the proportion of the slice that is yet to be traded.

\subsection{Autocorrelation analysis of the intraday trading volume}\label{sec:autocorrelation}
\begin{figure}[!htbp]
    \centering
    \includegraphics[width=0.63\linewidth]{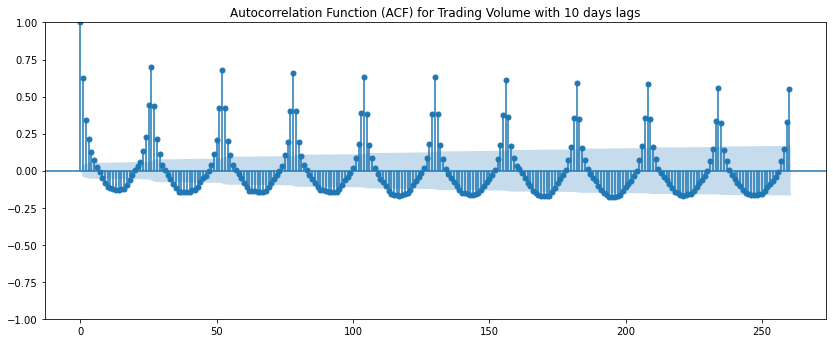}
    \caption{\textbf{ACF plotting of the intraday volume S\&P index}. The S\&P index is constructed using a weighted methodology, with weights assigned to each stock based on the number of trades of each stock in the respective bin. }
    \label{fig:acf}
\end{figure}

\Cref{fig:acf} shows the autocorrelation analysis of the intraday trading volume, the pattern of which indirectly corroborates the U-shape of the intraday volume.

\subsection{Summary of top contributing features in model prediction}
\begin{table}[!hbpt]
    \centering
    \caption{\textbf{Explanation of the Top Important Predictors}.}
    \begin{tabular}{ll}
    \toprule
    \textbf{Feature} & \textbf{Explanation} \\
    \midrule
    \texttt{eta} &  Daily component of volume \\
    \texttt{seas} & Intraday periodic component of volume \\
    \texttt{mu} & Intraday non-periodic component of volume  \\
    \texttt{x} & Multiplication of eta, seas, and mu,\\
     & for more details please refer to \eqref{eq:cmem} \\
    \texttt{ntn} & Trading amount, i.e. notional value of trades in USD \\
    \texttt{ntn\_8} & Trading amount in past 8 bins, i.e. past 2 hours.\\
    \texttt{daily\_qty} & Number of shares of orders in past day \\
    \texttt{intraday\_qty} & Number of shares of orders in the same interval \\
    & of the past day (open, mid-day, or close) \\
    \texttt{daily\_volBuyQty} & Daily number of shares in buy orders\\
    \texttt{daily\_volSellQty} & Daily number of shares in sell orders\\
    \texttt{nrTrades} & Number of trades.\\
    \texttt{volBuyQty} & Number of shares in buy orders .\\
    \texttt{volSellQty} & Number of shares in sell orders. \\
    \texttt{volBuyNotional} & Buy notional value in USD \\
    \texttt{volBuyNotional\_2} &  Buy notional value in USD\\
    & in past 2 bins, i.e. past half an hour.\\
    \bottomrule
    \end{tabular}
    \label{tab:feature_importance_explain}
    \end{table}
    
\restoregeometry

\subsection{Ablation study on neural network architectures}

We also investigate what part of the neural network architecture contributes the most to the observed improved performance. Under the SAM setting, we compare the model performance with the MLP+LSTM structure and the CNN+LSTM structure.

\begin{table}[!htbp]
    \centering
    \begin{tabular}{c c c c}
        \toprule
         & \multicolumn{2}{c}{$R^2$} \\ 
        \cmidrule(lr){2-3}
        Model Architecture & Mean & Std \\ 
        \midrule
        MLP+LSTM & 0.521 & 0.075  \\ 
        CNN+LSTM & 0.566 & 0.071 \\ 
        \bottomrule
    \end{tabular}
    \caption{Comparison of model performance of two network structures.}
    \label{tab:ablation}
\end{table}

From \Cref{tab:ablation}, it is evident that the DeepLOB structure proposed by \cite{deeplob}, incorporating CNN+LSTM, outperforms the MLP+LSTM configuration. This superior performance, with an out-of-sample $R^2$ of 0.420 for CNN+LSTM compared to 0.408 for MLP+LSTM, indicates that the DeepLOB structure more effectively extracts latent features and dependencies from the network input, thereby enhancing overall network performance. It is also noteworthy that early stopping is employed in these experiments to optimize the out-of-sample performance of the model. The efficacy of early stopping is further corroborated by \cite{Kolm2021DeepOF}, who employed the same network structure and limit order book data as input, but with a different forecasting target -- our target being trading volume, whereas theirs was order flow imbalance.

\subsection{Correlation matrix comparison in the CAM setting}  \label{sec:comparison_CAM}

\begin{figure}[!htbp]
    \centering
    \includegraphics[width=1\linewidth]{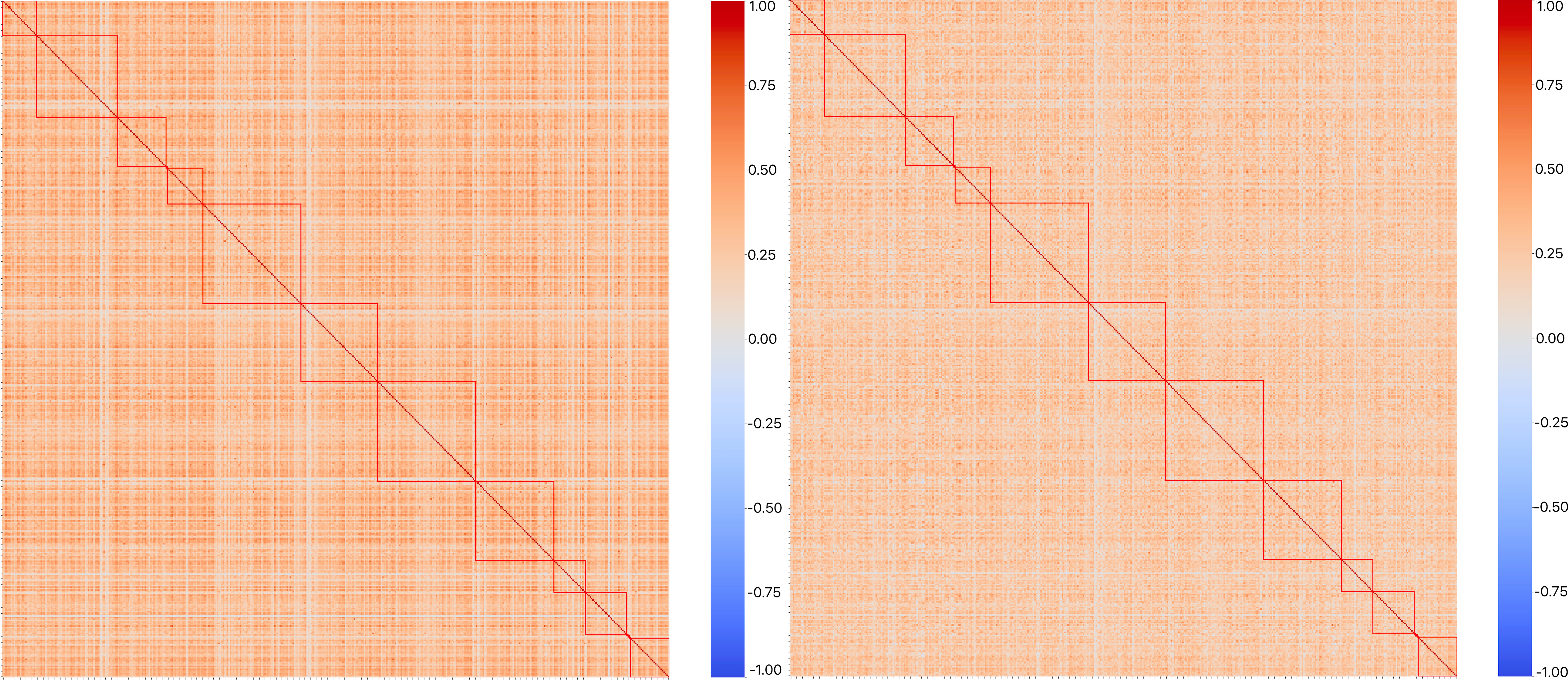}
    \caption{\textbf{Comparison of correlation matrix heatmaps}. The left heatmap is the correlation matrix computed based on trading volume data; the right heatmap is the correlation matrix computed based on the features for forecasting data. The red rectangles indicate correlations within each of the industry sectors.}
    \label{fig:heatmaps}
\end{figure}

We are also interested in assessing whether it is necessary to establish two distinct settings for computing the correlation matrix, as mentioned in \Cref{sec:cam_detials}. \Cref{fig:heatmaps} presents heatmaps of the two correlation matrices under the CAM setting, revealing distinct visual patterns between them. While the left heatmap based on trading volume data shows more pronounced correlations indicated by the greater number of red cells, both matrices share similar structural characteristics in terms of how correlations are distributed. Additionally, there are no clear diagonal blocks in either correlation matrix heatmap, which would have indicated strong within-sector correlations. This finding suggests that while visual differences exist between the correlation matrices derived from different data sources, the practical impact on clustering outcomes may be limited, suggesting that either method could be effective for constructing correlation matrices in the CAM framework. However, further investigation would be beneficial to fully understand the implications of these different correlation structures on forecasting performance.

\subsection{Detailed analysis of the clustered asset model} \label{sec:cluster_asset_model}

\begin{table}[!ht]
    \centering
    \caption{\textbf{Cumulative Sums and Number of Clusters Analysis}. This table presents the cumulative sums and cluster number analysis results from the trading volume cluster model, where \textbf{EVR Cumsum} represents the cumulative sum of the Explained Variance Ratio (EVR) after we apply PCA to the correlation matrix. The \textbf{Number of Clusters} represents the count of clusters selected when employing K-Means++ following PCA.} 
    \label{tab:clustered_table_v2}
    \begin{tabular}{cccc}
    \toprule
     & & \multicolumn{2}{c}{Volume Cluster Model} \\
    \cmidrule(lr){3-4}
    Analysis Type & Clusters Num & EVR Cumsum & \(R^2\) \\
    
    \midrule
                   & 10 & 0.80  & \textbf{0.596} \\
    Cumulative     & 10 & 0.99  & 0.596 \\
    Sum Analysis   & 10 & 0.999 & 0.595 \\
                   & 10 & 1.00  & 0.595 \\
    \midrule
    Number of   & 5  & 0.80  & 0.596 \\
    Clusters Analysis     & 10 & 0.80  & 0.596 \\
                 & 20 & 0.80  & 0.597 \\
                 & 50 & 0.80  & \textbf{0.600} \\
    \bottomrule
    \end{tabular}
    \end{table}

    From Table \ref{tab:clustered_table_v2}, we observe that for the features clustering model, the smaller cumulative sum yields better performance, suggesting that dimensionality reduction is beneficial for predictions.  Meanwhile, we notice that under the same PCA cumulative sum, smaller clusters exhibit better predictive performance, and the grouping of 50 stocks performs the poorest. We speculate that this is due to the commonality among trading volumes of the stocks, which aids in enhancing prediction accuracy.
    
    Moreover, by comparing the CAM and UAM models, we conclude that CAM performs better than UAM, which denotes that only the commonality among similar or neighboring stocks contributes significantly to improving predictions. The reason is that before clustering, the information during prediction is convoluted, and valuable intra-cluster information that could aid in predictions is not effectively extracted.

\section{Design of VWAP Strategy using Trading Volume Forecasts}

\label{sec:application_details} 

In Section \ref{sec:application}, we define the trading task as executing a volume equal to $P=1\%$ of the total trading volume  per security per trading day. This setting is common in scheduling tasks and is desired to have a minimal price impact on the dynamics of the limit order book. The model used in \Cref{sec:application} to be compared with the benchmark is the best-performing model proposed in \Cref{sec:CNN_LSTM}. For the experimental setup of quantifying the tracking error of the VWAP replication, we employed Level 2 order book data, assuming a direction for buying or selling (in our setting, we choose the selling direction), and then executing trades at the counterparty's best aggressive price, which is the first-level best price. Another approach to conducting experiments involved using the last price of each bin as the final transaction price, as proposed in \cite{kf_cmem2016}. Both approaches rely on past trading prices as a proxy for the price of their executed orders.

\section{Analysis on Order Flow Imbalance Predictors}\label{appendix:new_predictors}

In this section, we provide further investigation of the family of order flow imbalance \texttt{ofi} predictors (i.e., \{\texttt{ofi\_0}, \dots, \texttt{ofi\_9}, \texttt{best\_level}, \texttt{ofi\_cont}\}). Our goal is to understand and quantify how these additional variables influence intraday volume forecasting, and to provide a brief explanation of why signed order flow information (direction-based \texttt{ofi}) appears to outperform magnitude-only (\(\lvert \texttt{ofi}\rvert\)) variants in certain experiments.

\subsection{Definition of new predictors}
We start by providing three definitions of the \texttt{ofi}, followed by its signed and absolute variants. To establish a clear foundation for these predictors, we first define the concept of order flow, which serves as the basis for constructing the subsequent metrics.

Building on the definition provided in \cite{lob_definition}, we describe \textbf{\texttt{order flow}} as follows: over the interval $(t-h, t]$, all updates to the order book are indexed by $n$. For a given stock $i$ and two consecutive LOB states observed at $n-1$ and $n$, the bid-side order flows $(\text{OF}_{i,n}^{m,b})$ and ask-side order flows $(\text{OF}_{i,n}^{m,a})$ at depth level $m$ are computed as

\begin{align*}
\text{OF}_{i,n}^{m,b} &= \begin{cases} 
    q_{i,n}^{m,b}, & \text{if } P_{i,n}^{m,b} > P_{i,n-1}^{m,b} \\
    q_{i,n}^{m,b} - q_{i,n-1}^{m,b}, & \text{if } P_{i,n}^{m,b} = P_{i,n-1}^{m,b} \\
    -q_{i,n}^{m,b}, & \text{if } P_{i,n}^{m,b} < P_{i,n-1}^{m,b}
\end{cases} \\\\
\text{OF}_{i,n}^{m,a} &= \begin{cases} 
    -q_{i,n}^{m,a}, & \text{if } P_{i,n}^{m,a} > P_{i,n-1}^{m,a} \\
    q_{i,n}^{m,a} - q_{i,n-1}^{m,a}, & \text{if } P_{i,n}^{m,a} = P_{i,n-1}^{m,a} \\
    q_{i,n}^{m,a}, & \text{if } P_{i,n}^{m,a} < P_{i,n-1}^{m,a}
\end{cases}
\end{align*}

\begin{enumerate}
    \item \textbf{\texttt{ofi\_cont}}:  
    As defined in \cite{cont_ofi}, the Order Flow Imbalance (OFI) at the best price level is expressed as
    \begin{equation}
    \text{OFI}_{i,t}^{1,h} = L_{i,h}^{1,b} - C_{i,h}^{1,b} - M_{i,h}^{1,b} - L_{i,h}^{1,a} + C_{i,h}^{1,a} - M_{i,h}^{1,a}, 
    \end{equation}

\noindent where
    \begin{itemize}
        \item $L_{i,h}^{1,b}$ represents the total volume of new buy limit orders placed at the best bid during $(t-h, t]$,
        \item $C_{i,h}^{1,b}$ represents the total volume of buy limit orders canceled from the best bid during $(t-h, t]$,
        \item $M_{i,h}^{1,b}$ represents the total volume of marketable buy orders executed against the best ask during $(t-h, t]$,
        \item Similarly, $L_{i,h}^{1,a}, C_{i,h}^{1,a}, M_{i,h}^{1,a}$ denote the respective volumes for sell orders at the best ask.
    \end{itemize}

    \item \textbf{\texttt{best\_level}}:  
    The best-level OFI aggregates the net order flow at the best bid and ask prices over a given time interval, capturing imbalances at the top level of the order book (\cite{lob_definition, cont_ofi, deeper_level_ofi}). It is defined as
    \begin{equation}
    \text{OFI}_{i,t}^{1,h} := \sum_{n=N(t-h)+1}^{N(t)} \left( \text{OF}_{i,n}^{1,b} - \text{OF}_{i,n}^{1,a} \right).
    \end{equation}

    In this definition, $N(t-h)+1$ and $N(t)$ refer to the indices of the first and last order book events within the interval $(t-h, t]$, respectively.
    
    \item \textbf{\texttt{ofi\_n}}:  
    For a given price level $n$ (e.g., within the top 10 levels), the deeper-level OFI quantifies the net order flow on both the buy and sell sides, adjusted by the average order book depth. As described in \cite{lob_definition, xu_multi-level_2018, deeper_level_ofi}, this metric extends beyond the best price level, providing a comprehensive view of order flow dynamics across multiple levels.

    The OFI at level $n$ is defined as
    \[
    \text{OFI}_{i,t}^{n,h} := \sum_{k=N(t-h)+1}^{N(t)} \left( \text{OF}_{i,k}^{n,b} - \text{OF}_{i,k}^{n,a} \right). 
    \]

    To account for variations in order book depth, the scaled deeper-level OFI is expressed as
    \[
    \text{ofi}_{i,t}^{n,h} = \frac{\text{OFI}_{i,t}^{n,h}}{Q_{i,t}^{M,h}},
    \]
    where:
    \[
    Q_{i,t}^{M,h} = \frac{1}{M} \sum_{m=1}^M \frac{1}{2\Delta N(t)} \sum_{k=N(t-h)+1}^{N(t)} \left( q_{i,k}^{m,b} + q_{i,k}^{m,a} \right)
    \]
    represents the average depth across the first $M$ levels.

    Additional details include:
    \begin{itemize}
        \item $\text{OF}_{i,k}^{n,b}$ denotes the net change in buy-side order volumes at level $n$ between events $k-1$ and $k$, including new limit orders, cancellations, and modifications.
        \item $\text{OF}_{i,k}^{n,a}$ similarly denotes the net change on the sell side.
        \item $\Delta N(t)$ is the total number of order book events during $(t-h, t]$.
        \item $N(t)$ and $N(t-h)+1$ index the last and first events in the interval, respectively.
    \end{itemize}

    For practical purposes, we often consider the top $M = 10$ levels of the order book and represent the multi-level OFI as
    \[
    \text{ofi}_{i,t}^{(h)} = \left( \text{ofi}_{i,t}^{1,h}, \dots, \text{ofi}_{i,t}^{10,h} \right)^T.
    \]

    Net buying activity at level $n$ is indicated by a positive $\mathrm{ofi}_{i,t}^{n,h}$, while a negative value denotes net selling activity.

\end{enumerate}

\paragraph{Signed \texttt{ofi}.} 
The predictor \(\texttt{ofi\_n}\) computes signed net order flow at the \(n\)-th price level (scaled by an average depth). A positive \(\texttt{ofi\_n}\) indicates net buying, while a negative value indicates net selling.

\paragraph{\texttt{best\_level} and \texttt{ofi\_cont}.} 
\(\texttt{best\_level}\) accumulates net order flow exclusively at the top level, while \(\texttt{ofi\_cont}\) extends this notion across multiple top levels with finer-grained order breakdowns.

\paragraph{\(\lvert\texttt{ofi}\rvert\).} 
We take the absolute value of the OFI by removing the sign, denoted as $|\texttt{ofi}|$. This transformation omits directional buy/sell information and focuses on the magnitude of order flow.

\subsection{Experimental setup and sample results}

We incorporate the above predictors into our baseline feature sets (e.g., log-transformed volumes, rolling aggregated features, CMEM-derived components), and evaluate and compare out-of-sample \(R^2\).

\begin{table}[h]
\centering
\caption{Out-of-sample \(R^2\) (\(\text{mean}, \text{std}\)) for Models Incorporating \texttt{ofi}. 
Columns \texttt{Diff} represent the change in \(R^2\) compared to a baseline model (same linear or nonlinear type) without these predictors. \textbf{Linear} corresponds to a Ridge model, while \textbf{Nonlinear} corresponds to an XGB model. The bold typeface indicates the best performance within each panel. The first panel compares different variants of \texttt{ofi} (direction-based vs.\ absolute-based). The second panel shows how the three direction-based components in the \texttt{ofi} family (\texttt{ofi\_n}, \texttt{best\_level}, \texttt{ofi\_cont}) perform when included in the model.}
\label{tab:appendix_ofi_new}
\begin{tabular}{lcccccc}
\toprule
{\textbf{Predictors}} & \multicolumn{3}{c}{\textbf{Linear}} & \multicolumn{3}{c}{\textbf{Nonlinear}} \\
\cmidrule(lr){2-4}\cmidrule(lr){5-7}
 & \textbf{Mean} & \textbf{Std} & \textbf{Diff} & \textbf{Mean} & \textbf{Std} & \textbf{Diff} \\
\midrule
\texttt{ofi} (direction-based) & 0.494 & 0.060 & +0.003 & 0.544 & 0.079 & +0.012 \\
\texttt{ofi} (absolute-based)  & \textbf{0.494} & \textbf{0.060} & \textbf{+0.003} & \textbf{0.560} & \textbf{0.074} & \textbf{+0.028}  \\
\midrule
\texttt{ofi\_n}               & 0.492    & 0.059    & +0.001     & 0.538    & 0.079    & +0.006     \\
\texttt{best\_level}            & \textbf{0.493}   & \textbf{0.060}   & \textbf{+0.002}    & \textbf{0.543}   & \textbf{0.079}   & \textbf{+0.011}     \\
\texttt{ofi\_cont}              & 0.492   & 0.060    & +0.001     & 0.537    & 0.079    & +0.005     \\
\bottomrule
\end{tabular}
\end{table}

Table~\ref{tab:appendix_ofi_new} reveals that, in the first panel, \(\lvert \texttt{ofi}\rvert\) (absolute-based) slightly outperforms the direction-based approach in our nonlinear setting, suggesting that magnitude-based features can sometimes capture trading intensity more robustly. In the second panel, we compare how each direction-based \texttt{ofi} component (\texttt{ofi\_n}, \texttt{best\_level}, and \texttt{ofi\_cont}) improves forecast performance relative to the baseline. The \texttt{best\_level} performs the best, meaning that order flow information at the top price level contains particularly valuable signals for volume prediction. This superior performance suggests that activity at the best bid and ask prices, where most of the executable liquidity resides and where price discovery primarily occurs, provides the most relevant information for forecasting trading volume. The improved performance of \texttt{best\_level} ($+0.002$ in linear models, and $+0.011$ in nonlinear models) compared to \texttt{ofi\_n} and \texttt{ofi\_cont} indicates that the concentration of order flow at the most competitive prices at the top of the book more directly influences subsequent trading activity, compared to deeper levels in the order book or more granular order breakdowns. This aligns with market microstructure theory, which posits that the best price levels attract the most attention from market participants and therefore contain the most informative signals about future trading intentions.

\subsection{Discussion: why \(\lvert \texttt{ofi}\rvert\) often outperforms direction-based \texttt{ofi}}   

One possible explanation for the outperformance of \(\lvert \texttt{ofi}\rvert\)  is that when directional signals (indicating buying or selling pressure) are hard for the model to learn, possibly due to limited sample size or high noise, the absolute measure of order flow $\lvert \texttt{ofi}\rvert$ can be more robust.
Although direction-based \(\texttt{ofi}\) indeed contains richer information, if the data and model are insufficient to distinguish reliable directional cues, then focusing purely on magnitude may lead to better and more stable predictive performance. In volatile regimes, absolute-based features can consistently capture the scale of trading pressure, while direction-based features may be more sensitive to noise and require larger datasets or additional supporting features.

\end{document}